\documentclass{jfm}
\usepackage[font=small,labelfont=bf]{caption}
\usepackage{subcaption}
\usepackage{graphicx}
\usepackage{newtxtext}
\usepackage{newtxmath}
\usepackage{natbib}
\usepackage{rotating}
\usepackage{subcaption}
\usepackage{amsmath}
\usepackage{amssymb}
\usepackage{csquotes}
\usepackage{graphicx}
\usepackage{subcaption}
\usepackage{caption}
\usepackage{hyperref}
\usepackage{amssymb}
\hypersetup{
    colorlinks = true,
    urlcolor   = blue,
    citecolor  = black,
}

\newcommand{\RomanNumeralCaps}[1]
\linenumbers

\title{Ultimate regime in Rayleigh-Darcy Convection.}

\author{Garima Varshney\aff{1}, Anikesh Pal\aff{1} \corresp{\email{pala@iitk.ac.in}}, and Narsimha Reddy Rapaka\aff{2,3} }

\affiliation{\aff{}\aff{1}Department of Mechanical Engineering, Indian Institute of Technology, Kanpur 208016, India.
\aff{2} Department of Mathematics, College of Computing and Mathematical Sciences, Khalifa University of Science and Technology, Abu Dhabi, United Arab Emirates.
\aff{3} Emirates Nuclear Technology Center, Khalifa University of Science and Technology, Abu Dhabi, United Arab Emirates.
}

\begin{document}
\maketitle

\begin{abstract}
Direct numerical simulations (DNS) of Rayleigh-Darcy convection in a 3D porous domain are performed at Rayleigh numbers $\in [10^3, 10^6]$ to investigate the heat-transfer scaling, thermal boundary-layer dynamics, and flow-structure evolution in the previously unexplored ultimate regime. The Nusselt number exhibits an approximately linear dependence on the Rayleigh number throughout the investigated range. However, a distinct change in slope is observed at $Ra \approx 4\times10^5$, indicating the onset of the ultimate regime. For $Ra \leq 2.5\times10^5$, the obtained scaling is $\sim 6.25\%$ lower than that reported by \cite{hewitt2014high} while for $Ra \geq 4\times10^5$ our results are within $\sim 1.24\%$ of the extrapolated ultimate-regime prediction of \cite{pirozzoli2021towards}, demonstrating excellent agreement with the literature. Analysis of the thermal structure reveals the formation of near-wall protoplumes that merge into large-scale columnar megaplumes. With an increase in $Ra$, the size of the protoplumes decreases, whereas the numbers increase, thus enhancing boundary-layer convection and heat transport. The thermal boundary-layer thickness, defined as the variance of the temperature field, scales as $\sim Ra^{-1}$ and $\sim Nu^{-1}$, corroborating the persistence of linear heat-transfer scaling in the ultimate regime. The thermal dissipation is found to be increasingly shifting from the boundary layer to the bulk with increasing $Ra$, further indicating that the finer protoplumes efficiently transport the heat from the walls to the bulk. The flow structures are quantified using the dominant length scale in terms of the mean wavenumber ($\overline{k}$). It exhibits linear variation with $Ra$ for near-wall structures, with a higher slope for the ultimate regime ($\overline{k}\ \sim 0.0217Ra$ for $Ra\leq 2.5\times10^5$ and $\overline{k}\sim 0.023 Ra$ for $Ra \geq 4\times10^5$) signifying the finer protoplumes in the ultimate regime. At the mid-plane, the weaker scaling $\overline{k}\sim Ra^{0.489}$ suggests that the megaplumes also become finer with increasing $Ra$ in the ultimate regime, thus leading to efficient heat transport in the bulk.\\

\end{abstract}

\begin{keywords}

\end{keywords}
\section{Introduction}
\label{sec:headings}

Anthropogenic activities have significantly increased carbon dioxide (CO$_2$) emissions, which could potentially become a devastating issue: \enquote{Global Warming}. One feasible approach to mitigate its impact on the environment is to extract CO$_2$ from stationary sources such as the burning of fossil fuels and exhaust from industrial chimneys and sequester it beneath the Earth's surface, e.g., in depleted oil and gas reservoirs, saline aquifers, deep coal beds, and sea sediments. Of these, saline aquifers emerge as promising potential sites due to the highest storage capacity \citep{bruant2002safe} and long-term reliability of safe capture (\cite{liu2024determination}, \cite{metz2005ipcc}). When CO$_2$ is introduced into the aquifers at a depth of more than $800$m
it achieves its supercritical (or liquid-like) state (\cite{huppert2014fluid}). 
The supercritical CO$_2$ (scCO$_2$) is buoyant relative to the brine, therefore remaining atop the water and susceptible to leakage if the upper confining layer (caprock) breaks due to imbalancing of buoyancy, pressure, and gas injection pressure (\cite{rapaka2009onset}, \cite{emami2015convective}, \cite{Tsang2017}).  
A mixture boundary is established between scCO$_2$ and brine due to the partial solubility of scCO$_2$ in brine water. This mixing zone has a higher density $\sim$ 1\% than the adjacent brine \citep{rapaka2008non}, resulting in {\color{black}gravitational instability} that produces finger-like complex patterns during its descent, thus facilitating sequestration of CO$_2$ through convective dissolution \citep{riaz2006onset}.
To understand the different aspects of these instabilities, such as the onset and post-injection migration, and the time evolution of fluxes, extensive research has been carried out using stability analysis (\cite{farajzadeh2013empirical}, \cite{rapaka2008non} etc.), nonlinear simulations (\cite{hewitt2014high}, \cite{neufeld2010convective}, \cite{hewitt2020vigorous} etc.), and experiments (\cite{eckel2022spatial}, \cite{tsai2013density}, \cite{liyanage2019multidimensional} etc.) in two dimensional (2D) and three dimensional (3D) domains with porous media. These investigations were carried out on different types of porous media to investigate CO$_2$ sequestration (\cite{huppert2014fluid}, \cite{bashir2024comprehensive}), heat transport in volcanic systems \citep{hewitt2020vigorous}, 
geothermal energy extraction \citep{CHENG19791}, oil extraction \citep{varshney2023numerical}, formation of freckles
in alloy castings \citep{fowler1985formation} and other industrial applications (\cite{nield2006convection}. \\

Convection in such porous media is governed by Darcy's law and the advection-diffusion of density (temperature) or solute, and is modelled as Rayleigh-Darcy Convection (RDC). The only non-dimensional parameter encountered here is the Rayleigh-Darcy number $(Ra)$ resulting in the characterization of heat (or mass) flux in terms of the Nusselt number $(Nu)$ (or the Sherwood number $(Sh)$) (detailed discussion will be covered in section \ref{sec:section2.1}).
Depending on whether the flow is transient or steady, studies are conducted in two different ways: (\textit{a}) one-sided (transient), where the density-driven dissolution starts only at the upper boundary and evolves over time, (\textit{b}) two-sided convection, where the buoyancy source extends across two boundaries (hot and cold), ultimately transitioning to a steady state. Statistical steady-state analysis of the latter provides more insight in such convection problems than the transient case (\cite{Hewitt_Neufeld_Lister_2013_}, \cite{hewitt2020vigorous}). \\

\cite{horton1945convection} and \cite{lapwood1948convection} were the pioneers in the convection of isotropic porous media, focusing on one-sided convection. By extending the classical theory of thermal convection to permeable domains, \cite{horton1945convection} provided one of the earliest quantitative frameworks for evaluating the stability of a fluid-saturated porous layer subjected to a vertical temperature gradient. \cite{lapwood1948convection} formalized this threshold using the Rayleigh–Darcy number, a dimensionless parameter that remains central to the study of porous media convection. 
Building upon this classical foundation, later studies have extended the theory to more complex and practical scenarios. \cite{riaz2006onset} investigated the onset and development of diffusive boundary layers using linear stability analysis, incorporating a self-similar diffusive base state. They extended their analysis by employing quasi-steady-state approximation for accurate solutions and validated their predictions through high-resolution DNS. \cite{rapaka2008non} predicted the maximum growth rate for any perturbation at a given time using non-modal stability analysis, along with anisotropic and layered heterogeneous media. They exhibited a notable enhancement over prior linear stability results by emphasizing the prominence of nonlinear components. \cite{rapaka2009onset} further used their theory to predict the growth of perturbations in anisotropic and layered heterogeneous porous materials. Theoretical modeling and high-resolution numerical simulations of one-sided convection in porous media performed by \cite{hidalgo2012scaling} reported that in the regime of high Rayleigh numbers, the mean scalar dissipation rate governs the dissolution flux, which remains constant and independent of the Rayleigh numbers. This result challenges earlier expectations that convective flux would continue to scale with increasing buoyancy forces. \cite{tsai2013density} conducted the experiments and examined global convective transport for the range of $10^4 \leq Ra \leq 10^5$ for Hele-Shaw cells and porous domains. They reported the scaling as $Nu=ARa^\beta$ with A = 0.037 and $\beta$ = 0.84, which is consistent with the established findings of $\beta \approx$ 4/5 \citep{neufeld2010convective}. \cite{ching2017convective} explored the convective mixing during CO$_2$ dissolution using Hele-Shaw cells for the range $2 \times 10^4 \leq Ra \leq 8.26 \times 10^5$ and confirmed that the $Nu-Ra$ relationship is largely independent of $Ra$. This suggests that once convection is fully developed, the efficiency of mixing is governed by the plume structure rather than further increases in buoyancy forces. 
Recently, \cite{guo2021novel} experimentally investigated the spatial and temporal evolution of quasi-2D convection to determine the critical parameters ($Ra_c$ and $t_c$) for the onset of convection and its dissolution. Their experiments demonstrated that variations in permeability and initial solute concentration significantly influence finger formation and mixing dynamics and established the empirical linear correlation based on the top boundary mass flux.\\

\cite{fu2013pattern} performed high-resolution 3D simulations over the range of 1600 $\leq Ra \leq$ 6400 to study the pattern formation aspects of convective mixing in porous media. They reported the emergence of cellular network structures in the diffusive boundary layer. They further used the statistics of these cellular networks to identify various regimes of finger coarsening over time, the existence of a non-equilibrium stationary state, and a universal scaling of 3D convective mixing.  A high-resolution MRI system was employed by \cite{teng2017quantifying} to experimentally investigate convection in a porous packed vertical tube. The relation between wave number, wavelength, onset time, and mixing time as functions of $Ra$ is examined to study the finger lifecycle from inception to propagation, coalescence, and the formation of further fingers. With MEG-water as the model fluid, \cite{liyanage2019multidimensional} conducted non-invasive convective dissolution experiments using X-ray computed tomography in a uniform 3D porous medium and found the evolution of the mixing process from pure diffusion to convection-dominated and eventual shutdown. They report a linear relationship between the non-dimensional convective flux, the Sherwood number ($Sh$), and the Rayleigh number as $Sh=0.0025Ra$ for $Ra < 5000 $. X-ray CT imaging was also used by \cite{eckel2022spatial} to investigate the evolution of the convective mixing process for different regimes in 3D homogeneous porous media in the range of $Ra$= 3000–55000 and proposed a power-law relationship $Sh = 0.18Ra^{0.77}$.\\

Convection from both boundaries is also modelled as Rayleigh-Darcy convection by constraining a fluid-saturated porous zone between hot and cold plates. \cite{otero2004high} conducted two-dimensional DNS to investigate convection modes and assess heat transfer for $Ra = 4\pi^2$ to $10^4$. Their results revealed two distinct dynamical regimes: a temporally periodic (chaotic) regime, marked by a dominant pair of rolls formed at the onset, with heat transfer following the classical scaling law, $ Nu\sim Ra$, and a turbulence regime, marked by the breakdown of long-range spatial coherence and detachment of thermal plumes, resulting in a modified scaling behavior $Nu\sim Ra^{0.9}$. The authors suggested that the deviation from classical scaling might be attributed to transient effects or the absence of logarithmic corrections in the analyzed range. \cite{neufeld2010convective} combined numerical simulations with laboratory experiments using solutions of methanol and ethylene‐glycol (MEG) mixing with water to study the convective behavior of CO$_2$-enriched brine. They reported a non-linear relationship between the convective flux scale and the Rayleigh number as $Sh\propto Ra^{4/5}$. Further, they suggested that during the convection process, a diffuse boundary layer at the interface between CO$_2$ (or MEG) and water forms that becomes unstable to saturated fingers, which are born, swept aside, and coalesce in a mixing region before descending into the far field as long‐lived plumes. \cite{hewitt2012ultimate} also performed 2D Rayleigh-B\'enard (RB) convection in a porous medium for $Ra \leq 4 \times 10^4$, and found the linear classical scaling for the Nusselt number, $Nu \sim Ra$, is attained asymptotically. They also identified a distinct transition at $Ra \approx 1300$, marked by a sharp discontinuity in $Nu$, beyond which a scaling of $Nu \sim Ra^{0.95 \pm 0.01}$ similar to \cite{otero2004high} was observed. Their study also revealed the emergence of complex fishbone structures characterizing the temporal and lateral length scales associated with megaplumes and protoplumes. Additionally, a heat exchanger model was introduced to describe inner megaplume dynamics, leading to a horizontal wavenumber scaling of $k \sim Ra^{0.4}$. \cite{Hewitt_Neufeld_Lister_2013} numerically and theoretically examined the evolution of the dynamics and the convective flux during the shutdown regime for $Ra > O(10^3)$. They found that the evolution of flux in one-sided convection can be directly estimated from the convective flux in an RB cell of \cite{hewitt2012ultimate}.  They used this estimation of the flux to develop a box model that successfully predicts the dynamical structure of the flow in the shutdown regime and found a remarkable similarity to that in an RB cell. These ideas were extended to model convective systems
comprising two fluid layers, with an interface that is free to move, using a combination of simple theoretical
box models, high-resolution numerical simulations, and experiments, and it was reported that the interfacial deformation can dramatically enhance the convective flux. This convection process in a porous medium at high $Ra$ exhibits a quasi-steady columnar structure with a well-defined and $Ra$-dependent horizontal scale. The mechanism that controls this horizontal scale was unexplored till \cite{Hewitt_Neufeld_Lister_2013_} investigated the density-driven ‘heat-exchanger’ flow in a porous medium. They employed the Floquet theory for linear stability analysis to reveal that the flow is always unstable and further used DNS to show that the nonlinear evolution of the flow instability ultimately reduces the horizontal wavenumber of the background flow. \cite{wen2015structure} also performed spatial Floquet
analysis to reveal the existence of two types of instability for different aspect ratios at large $Ra$ in two-dimensional porous medium convection: a bulk
instability in which the most unstable disturbance spans the convective layer and a
wall instability in which the most unstable disturbance is strongly localised near the
hot and cold boundaries. They also carried out DNS to investigate the fully nonlinear flow evolution and suggested that the (long time) mean inter-plume spacing in statistically steady porous medium convection results from a balance between the competing effects of these two types of instability.  \\

Three-dimensional Rayleigh–Darcy convection remained largely unexplored till \cite{hewitt2014high} used high-resolution numerical simulations and reported a linear classical variation of $Nu$ with $Ra$ for $  1750 \le  Ra \le 2 \times 10^4$. Their $Nu$ was asymptotically approximately $40\%$ larger than that reported in the 2D investigation of \cite{Hewitt_Neufeld_Lister_2013}. They also found a resemblance of the interior flow with that of the $Ra\rightarrow \infty$ heat-exchanger model and reported the variation of the dominant horizontal wavenumber $k$ as $Ra^{0.5}$. This scaling is stronger than the two-dimensional scaling of $k \sim Ra^{0.4}$. This apparent difference in the dominant horizontal scale observed
in two-dimensional and three-dimensional DNS was explained by \cite{hewitt2017stability} using the stability of unbounded steady single-mode heat-
exchanger flow with a rectangular planform in a three-dimensional porous medium. More recently, \cite{hewitt2020vigorous} reviewed the theoretical, numerical, and experimental studies of the ‘two-sided’ and ‘one-sided’ systems and pointed out the challenges associated with high-$Ra$ porous convection. \cite{pirozzoli2021towards} overcame the numerical challenges and performed 3D simulations of Rayleigh–Darcy convection inside a cubic porous layer up to Rayleigh–Darcy number $Ra = 8 \times 10^4$. They reported that the Nusselt number variation at sufficiently high $Ra$ can be
approximately expressed as $Nu = 0.0081Ra + 0.067Ra^{0.61}$, with the error bar no larger than
1.5 \%. Furthermore, they provide an asymptotic scaling of $Nu = 0.0081Ra$ by extrapolating their prediction to the ultimate regime, which they expect to be reached at $Ra = 5 \times 10^5$. This scaling was almost 16\% less
than that inferred in the previous studies and exceeds by about $18\%$ of the Nusselt number
found in two-dimensional simulations. \cite{de2022strong} also performed simulations to study Rayleigh–Darcy convection in three-dimensional fluid-saturated porous media up to $Ra = 8 \times 10^4$, and demonstrated that the thickness of the thermal boundary layer scales inversely with the Nusselt number. They further reported the mean wavenumber scales as $Ra^{0.49}$ in the core of the domain, corroborating the theoretical and numerical predictions. However, near the domain boundaries, they find the mean wavenumber scaling as $Ra^{0.81}$, which is a deviation from the presumed linear behavior, concluding that the linear behavior of the mean wavenumber could only be observed in the ultimate regime where $Ra$ becomes $5 \times 10^5$ or higher. \\

It is apparent from the above discussion that convection in porous media continues to attract significant research attention owing to its critical role in a wide range of geophysical and engineering applications. 
These applications often involve extremely high Rayleigh numbers (e.g., $Ra\sim10^5-10^6$ and higher), as seen in real geological formations such as the Utsira Sand reservoir at Sleipner \citep{zhu2024transport, Hewitt_Neufeld_Lister_2013_}. Despite their importance, precise characterization of convection at such high $Ra$ remains challenging, and hence unexplored, both experimentally and numerically, due to the inherent complexity of subsurface systems and the resource-intensive nature of the simulations. \\

To the best of our knowledge, literature on Rayleigh-Darcy convection is limited to $Ra = 8 \times 10^4$. The $Nu$ in the expected ultimate region is estimated by extrapolating the scaling of $Nu$ upto $Ra = 8 \times 10^4$. In the present investigation, we perform 3D high-resolution simulations to study the Rayleigh-Darcy convection in fluid-saturated porous media for $Ra \in [10^3, 10^6]$. We aim to measure the $Nu$ and report an appropriate $Nu - Ra$ in the unexplored ultimate regime at $Ra \geq 5 \times 10^5$ relevant to the real-world geological situations.  Additionally, we want to explore whether the mean wavenumber near the boundaries scales linearly with $Ra$ in the expected ultimate regimes. These simulations will offer a more precise understanding of the transport behavior in such extreme conditions. \\

We present the problem formulation, governing equations, domain details, averaging procedures, numerical methodology, and validation in section \ref{sec:section2}. The results from these simulations, focusing on the heat transfer characteristics and flow dynamics, are discussed in section \ref{sec:section3}, particularly the scaling relationships using temperature statistics and morphological analysis for the distinct regimes. We conclude our findings in section \ref{section4}. \\

\section{Problem formulation}\label{sec:section2}

\subsection{Governing equations} \label{sec:section2.1}

We consider a 3D porous domain of constant porosity $\phi$ and uniform permeability, $\kappa$ as shown in Figure \ref{fig:schematic}. 
The domain is assumed to be saturated with a fluid having constant viscosity $\mu$. 
The horizontal dimensions in the $x$ and $y$ directions are chosen according to the $Ra$ to optimize the available computation resources. We have used the horizontal domain sizes, $L_x = L_y$, in the range $\in [0.015625, 1]$, while the vertical domain size, $L_z$, is kept constant at unity for all the cases. This selection of lower aspect ratios is motivated by the investigation of \cite{iyer2020classical}, where it was reported that the temperature and the fluctuating velocity fields are similar for aspect ratios ($L_x/L_z, L_y/L_z$) of $0.1$ and $1$ owing to a very thin boundary layer compared to the horizontal dimensions of the domain at high $Ra$. We also compute the dominant wavenumber ($\overline{k}$) (detailed discussion in section \ref{sec:section3.4}) near the wall and find that the domain lengths for all the cases are sufficient enough to accommodate the largest wavelengths, thus justifying the lower aspect ratio domains for the high $Ra$ cases. \\

The flow in a porous medium is inertia-free, incompressible, and is governed by Darcy's law \citep{pirozzoli2021towards}. 
According to Darcy's law, the flow velocity is linearly related to the pressure gradients and the buoyancy term. 
The temperature transport equation \ref{Eq:energy}, regulates the flow driven by thermal convection \citep{hewitt2020vigorous}, having temporal dependence and non-linearity within the system.  The $D$ represents the constant thermal diffusivity.
The temperature field ($\theta^*$) varies linearly in the vertical direction, with $\theta_{max}^*$ and $\theta_{min}^*$ being the temperatures of the bottom and top boundaries of the domain, respectively. The $\rho^*$ is the density of the fluid, which varies linearly with temperature as shown in equation \ref{Eq:density}, and $\Delta \rho^*$ is the density difference of the fluid at the top and bottom. Convection induced by density variations originates from differences in the temperature of the fluid (\cite{hewitt2020vigorous}) or the concentration of the solute (\cite{neufeld2010convective}), or both (\cite{hu2023effects}). The fluid in contact with the heated boundary gets lighter compared to the fluid nearer it. This creates unstable density variation with respect to initial position and moves vertically upwards owing to buoyancy. At the same time, colder fluid comes down, and a convection current is established. However, this process occurs only if the imposed heat flux is sufficiently strong to overcome the stabilizing effect of the density stratification. 
In Equation \ref{Eq:cont}-\ref{Eq:density}, all the variables are in dimensional form and are denoted by superscripts$^*$.

\begin{equation}\label{Eq:cont}
     \boldsymbol{\nabla} \cdot \boldsymbol{u^*} = 0,
\end{equation}
\begin{equation}\label{Eq:Darcy}
\boldsymbol{u^*}= - \frac{\kappa}{\mu} \left( \boldsymbol{\nabla}{p^*} - \rho^* g\hat{\boldsymbol{k}}\right),
\end{equation}

\begin{equation}\label{Eq:energy}
       \phi \left(\frac{\partial{\theta^*}}{\partial t^*}\right) +\boldsymbol{\nabla} \cdot (\boldsymbol{u^*}\theta^*  - \phi D {\boldsymbol{\nabla} \theta^*})=0,
\end{equation}

\begin{equation}\label{Eq:density}
       \rho^*(\theta^*)= \rho^*  (\theta_{min}^*)- \Delta \rho^* \left(\frac{\theta^* -\theta_{min}^*}{\theta_{max}^* -\theta_{min}^*}\right).
\end{equation}


\subsection{Dimensionless Equations} \label{sec:section2.2}

The equations (\ref{Eq:cont}, \ref{Eq:Darcy}, \ref{Eq:energy}) are non-dimensionalised using the following set of dimensionless variables:  $\theta$ (temperature), $t$ (time), $p$(pressure), and $\boldsymbol{u}$ (velocity), defined by Equations \ref{non-dim}\hyperlink{partE}{a}-\hyperlink{partE}{d} respectively (\cite{fu2013pattern}, \cite{de2022strong}), using buoyancy velocity (${V}$)  (Equation \ref{non-dim}\hyperlink{partE}{e}) and the characteristic length scale ($L_z$) of the domain. 
The nondimensionalized forms of the governing equations are \ref{Eq:cont_nondim}, \ref{Eq:Darcy_nondim} and \ref{Eq:energy_nondim}. Here, ${\boldsymbol{u}}$, $p$, and $\theta$ are computed at the cell center, while $\boldsymbol{U_i}$ is the contravariant velocity computed on the faces of the cells. 
The Rayleigh–Darcy number ($Ra$), as defined in equation \ref{Eq:Ra}, appears as the only dimensionless parameter in the system and represents the ratio of buoyancy to dissipative effects (\cite{hewitt2014high}). 
It is referred to as the Rayleigh number ($Ra$) in further discussions for simplicity. \\


\refstepcounter{equation}
$$
    \theta = \frac{\theta^*-\theta_{min}^*}{\theta_{max}^*-\theta_{min}^*}, 
    \quad t = \frac{t^*}{\phi L_z/ {V}}, 
    \quad    p = \frac{p^*}{\Delta \rho g L_z},   
    \quad \boldsymbol{u}= \frac{\boldsymbol{u}^*}{{V}},
    \quad {V} = \frac{ g \Delta \rho \kappa}{\mu} \eqno{(\theequation{\mathit{a}-\mathit{e}})}  \label{non-dim}
$$
\newcommand{\partE}{V^* = \frac{g \Delta \rho k}{\mu}} 
\hypertarget{partE}{}

\begin{equation}\label{Eq:cont_nondim}
    {\mathit{\boldsymbol{\nabla}}} \cdot {\mathit{\boldsymbol{u}}} =0
\end{equation}
\begin{equation}\label{Eq:Darcy_nondim}
    {\mathit{\boldsymbol{u}}}=-({\mathit{\boldsymbol{\nabla}}} p-\theta {\mathit{\hat{\boldsymbol{k}}}})
\end{equation}
\begin{equation}\label{Eq:energy_nondim}
       \frac{\partial{\theta}}{\partial t} + {\mathit{\boldsymbol{\nabla}}} \cdot ({\mathit{\boldsymbol{U_i}}} \theta -\frac{1}{Ra}{\mathit{\boldsymbol{\nabla}}}\theta) =0
\end{equation}
\begin{equation} \label{Eq:Ra}
       Ra=\frac{{\mathit{g}} \Delta \rho \kappa L_z}{\phi D \mu}=\frac{{V}L_z}{\phi D}
\end{equation}


\subsection{Problem setup and Averaging technique} \label{sec:section2.3}

Table \ref{tab:case details} lists the cases corresponding to different $Ra$, with other dimensional and computational details. We used non-uniform grid spacing in the vertical direction and ensured sufficient grid points within the thermal boundary layer ($\delta _{\theta}$) edge, defined as the peak location of the temperature variance. This is achieved by clustering the grid near the walls using a hyperbolic stretching function defined as \citep{naskar_2022a, naskar_2022b,naskar2025energy,singhp2023,singh2025evolution,mishra2026evolution} 
\begin{equation} \label{Eq:Stretch}
z(m) = \frac{\tanh \Bigl\{ rz \left( \frac{m - 1}{N_z} - \frac{1}{2} \right) \Bigr\}}{2 \tanh \left( \frac{rz}{2} \right)}.
\end{equation}

Here, $m$ denotes the grid index, $N_z$ is the total number of grids in the vertical direction, and $rz$ is the stretching factor that controls the degree of refinement near the boundaries. The temperature gradients become increasingly concentrated near the walls at high $Ra$ \citep{pirozzoli2021towards,hewitt2014high}. Therefore, to accurately capture these thermal gradients, proper grid clustering is ensured near the boundaries as shown in figure \ref{fig:GRID}. In wall-parallel directions ($x$ and $y$) uniform grids ($N_x=N_y =384$) are used for each case.\\

We use periodic boundary conditions in the horizontal directions, assuming that the quantities are statistically invariant in these directions. Therefore, the horizontal average of any variable $\mathbf{A}(x,y,z,t)$ can be computed as 

\begin{equation} \label{Eq:Horizon_avg}
\overline {\mathbf{A}(z,t)} = \frac{1}{L_x L_y}\int_{0}^{L_x} \int_{0}^{L_y} \mathbf{A} (x,y,z,t) dy dx.
\end{equation}

The top and bottom boundaries are subjected to no-slip, impermeable, and isothermal conditions for velocity and temperature, respectively. The dimensionless boundary condition at the bottom and top of the domain for velocity and temperature are 

\begin{subequations} \label{eq:bc}
\begin{align}
\theta(L_z=0) &= 1, \quad u_z(L_z=0) = 0, \\
\theta(L_z=1) &= 0, \quad u_z(L_z=1) = 0.
\end{align}
\end{subequations}

The key response parameter in the system is the Nusselt number ($Nu$), which is the ratio of convective to diffusive heat flux and is calculated as\\ 
 
 \begin{equation}\label{Eq:Nu}
Nu = -\biggl \langle \frac{0.5}{L_x L_y} \int_{0}^{L_y}\int_{0}^{L_x} \frac{\partial\theta}{\partial z} \bigg|_{z=0} + \frac{\partial\theta}{\partial z} \bigg|_{z=1} dx dy \biggl \rangle,
\end{equation}
and is listed for each case in Table \ref{tab:case details}. The $\langle \cdot \rangle$ represents the temporal averaging. \\ 
\subsection{Numerical methodology}\label{sec:section2.4}

\begin{figure}
    \begin{subfigure}{0.35\linewidth}
        \includegraphics[width=\linewidth]{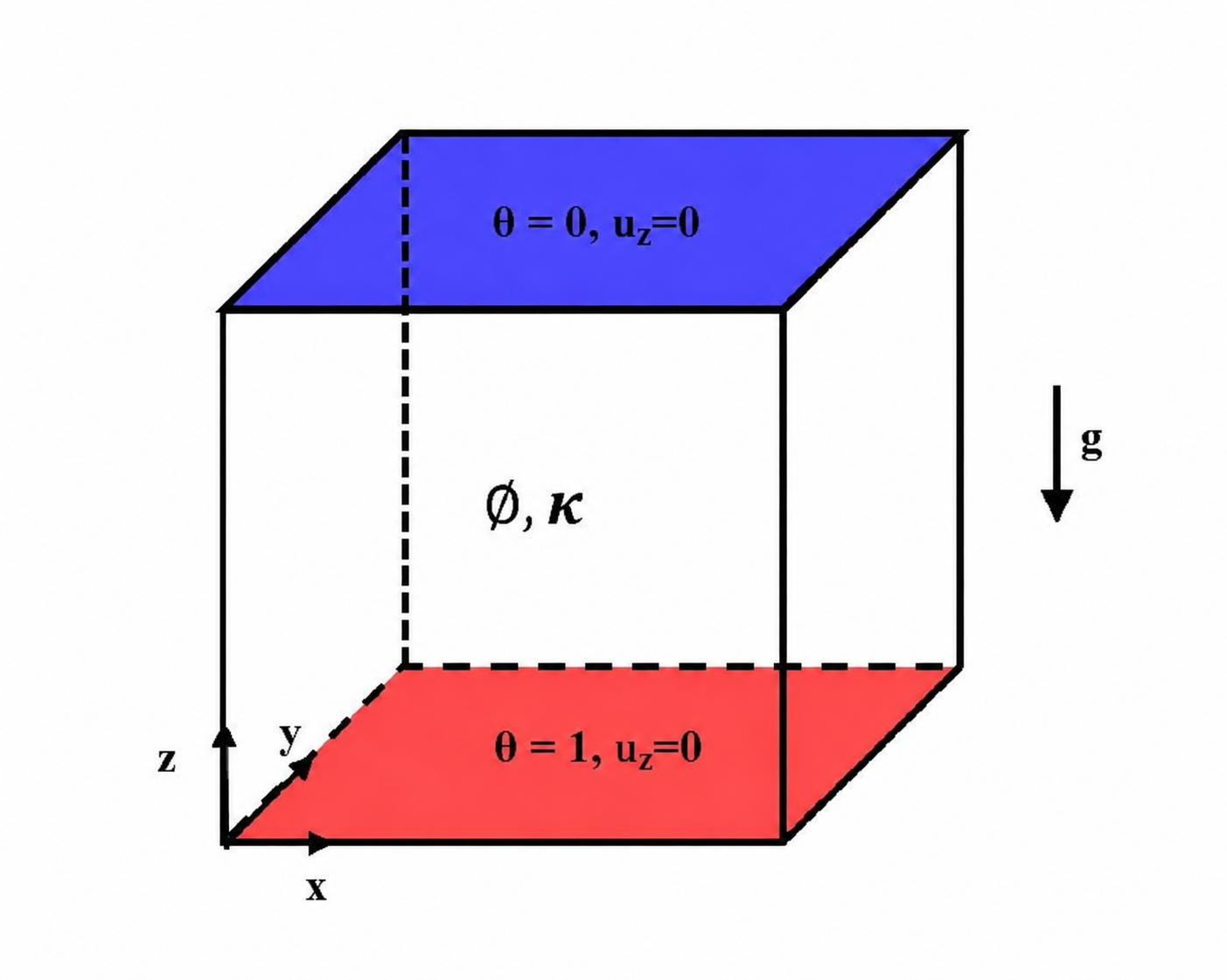}
        \caption{\label{fig:schematic}}
    \end{subfigure}
   \hfill
     \begin{subfigure}{0.35\linewidth}
         \includegraphics[width=\linewidth]{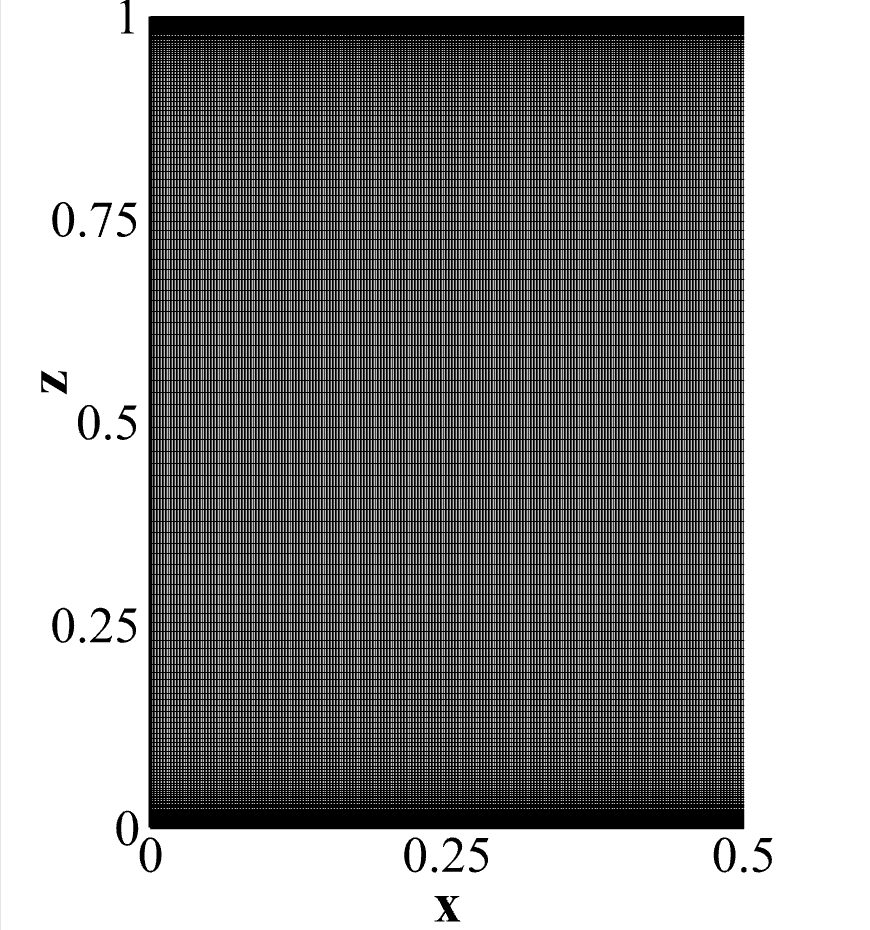}
         \caption{\label{fig:GRID}}
     \end{subfigure}
    \centering
    \hfill
    \begin{subfigure}{0.28\linewidth}
        \includegraphics[width=\linewidth]{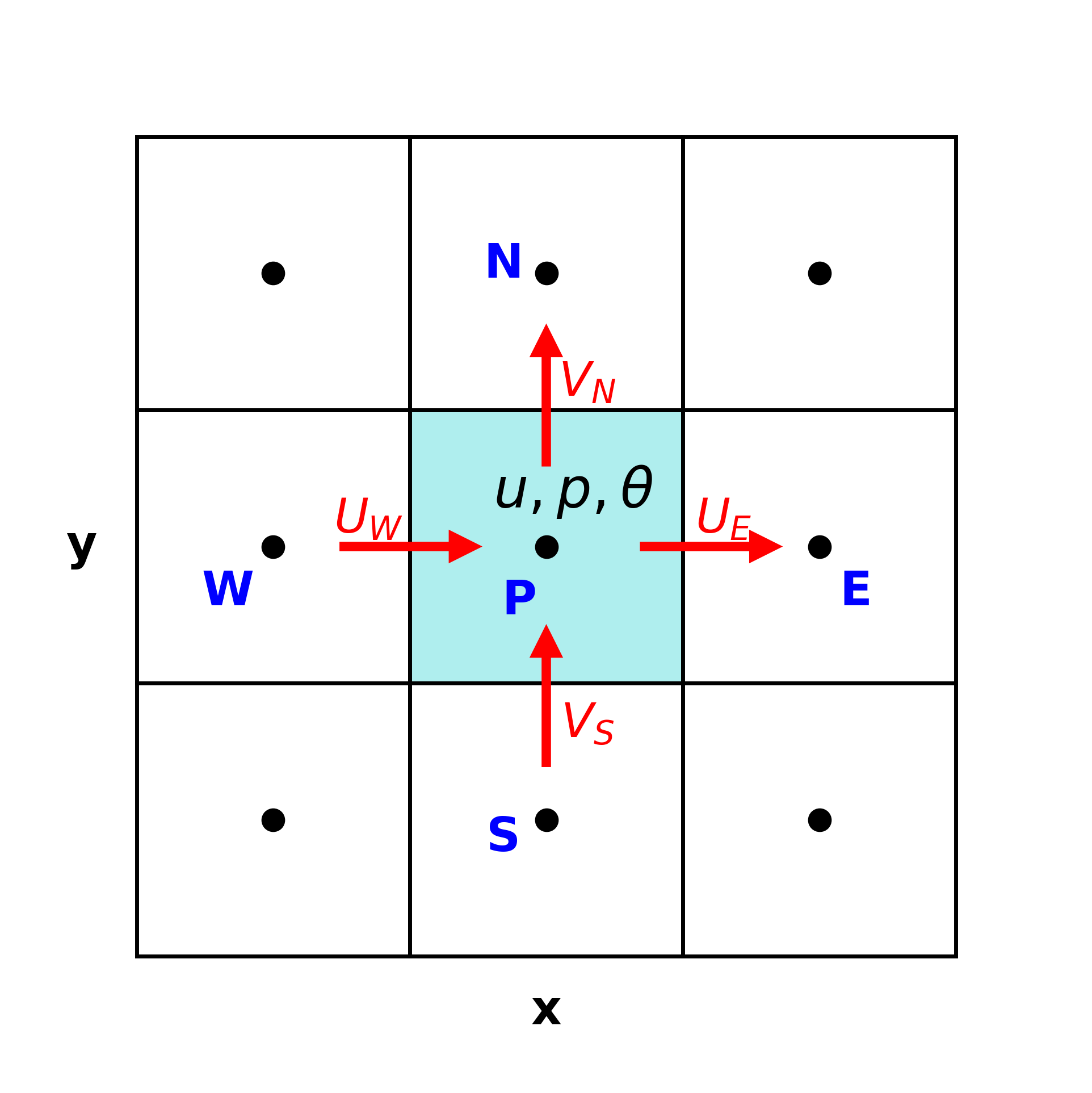}
        \caption{\label{fig:cell}}
    \end{subfigure}
    \vspace{0.5cm}
\captionsetup{justification=justified, singlelinecheck=false,width=\textwidth}
    \caption{\label{fig:1} (a) Schematic representation of the cuboidal porous domain  characterized by constant domain property $\phi,\kappa$, saturated with a mixture of $scCO_2$ and brine, featuring cold ($\theta =0$ at $L_z=1$) and warm boundaries ($\theta= 1$ at $L_z=0$) at the top (blue) and bottom (red), respectively. The domain length in the $z$ direction remains constant across all the cases, while optimized lengths are utilized for the horizontal dimensions.  The gravitational acceleration ${\boldsymbol{g}}$ is directed downward. (b) Collocated grid system, where velocity ($\boldsymbol{u}$), pressure ($p$) and temperature ($\theta$) are calculated at the cell center \enquote*{P} and contravariant velocity $\boldsymbol{U_i}$ is calculated at the face center using neighboring cells ($nb$=E, W, N, S, T, B). (c) Grid clustering near the wall. }
\end{figure}

The solution of equations \ref{Eq:cont_nondim}, \ref{Eq:Darcy_nondim}, and \ref{Eq:energy_nondim} is obtained using finite difference discretization in all three spatial directions using the collocated grid system. Each cell contains velocity ($\boldsymbol{u}$), pressure (p), and temperature ($\theta$) values, all defined at the cell center as shown in figure \ref{fig:cell}. The contravariant velocities ($\boldsymbol{U_i}$) are computed at the cell faces using the center values of the adjacent cells. These face-centered velocities are offset from the cell center by half of the cell width in each direction and are employed to advect fluxes originating from the cell center.
The advection and Poisson source terms utilize it to prevent the pressure-velocity decoupling. The solver utilizes a second-order central finite-difference scheme for spatial discretization, and a mixed RK3-ADI marching scheme {\color{black}\citep{rapaka2016immersed}} is used to advance the solution in time. 
The diffusion terms are advanced implicitly in all three spatial directions using the Alternating Direction Implicit (ADI) method, while an explicit low-storage Runge-Kutta-Wray3 (RKW3) time-marching scheme is used for other terms. For each RK3-ADI substep, a predictor-corrector algorithm is used to advance in time using the discretized Darcy equation along with the continuity equation. The velocity field is projected using the pressure and temperature of the preceding time step ($n$) using equation \ref{eq:ibm_u}. The contravariant velocities are then determined using the colocated velocity and the pressure of neighboring cells by the following equation \ref{eq:contra_u}. Here, overbar signifies a linear interpolation between the center of the cell \enquote*P and its neighboring cells (nb = E, W, N, S, T, B) separated by the corresponding face ($f$) and $\tilde u^*$ denotes the colocated velocity under zero pressure gradient determined by equation \ref{eq:predi_u}. The central difference schemes provide little numerical diffusion and show checkerboard pressure associated with pressure-velocity decoupling (\cite{ZANG199418}, \cite{rapaka2016immersed}). To mitigate this, a modified pressure gradient is added to $\boldsymbol{U_i}$ (equation \ref{eq:contra_u}), so that $\boldsymbol{U_i}$ on the face is calculated based on the pressure difference between the centers of the neighboring cells (P,nb) (\cite{rhie1983numerical}). 
The velocities are subsequently made divergence-free by solving the Poisson equation (Equation \ref{eq:divergence}) using the parallel multigrid iterative solver for pressure correction ($p'$) and applying corrections to the predicted velocity and pressure using equations \ref{eq:u_correct} and \ref{eq:p_correct} respectively. 
Here, the pressure gradient at the center of the cell (equation \ref{eq:u_correct}) and at the face of the cell (equation \ref{eq:U_correct}) ensures the coupling between velocity and pressure. The solver employs a domain decomposition method that utilizes the Message Passing Interface (MPI) for parallel processing to optimize the computational time and cost.
The time step is calculated on the basis of the Courant–Friedrichs–Lewy number for all the cases.

\begin{equation}\label{eq:ibm_u}
\boldsymbol{u} = -{\nabla p^n\big|_P}+ \theta^n\big|_P {\hat{\boldsymbol{k}}} 
\end{equation}

\begin{equation}\label{eq:predi_u}
{\boldsymbol{\tilde u^*}} = {\boldsymbol{\boldsymbol{u}}}+{\nabla p^n\big|_P}
\end{equation}

\begin{equation}\label{eq:contra_u}
{\boldsymbol{\tilde U^*_i}} = \overline{{\boldsymbol{\tilde u^*_P \Tilde{u_{nb}}}}}-{\nabla p^n\big|_f}
\end{equation}

\begin{equation}\label{eq:divergence}
\nabla \cdot \boldsymbol{U_i} = -{\nabla ^2 p'}
\end{equation}

\begin{equation}\label{eq:u_correct}
{\boldsymbol{u^{n+1}}} = {\boldsymbol{u}} +{\boldsymbol {\nabla}} p^{'}\big|_P +\theta^n\big|_P {\hat{\boldsymbol{k}}} 
\end{equation}

\begin{equation}\label{eq:U_correct}
{\boldsymbol{U^{n+1}}} = {\boldsymbol{U_i}} +{\boldsymbol {\nabla}} p^{'}\big|_f
\end{equation}
\begin{equation}\label{eq:p_correct}
p^{n+1} = p^n + p'
\end{equation}

\subsection{Validation} 


The solver has been extensively validated by \cite{rapaka2016immersed} for both stratified and unstratified turbulent flows, including the canonical unstratified turbulent channel flow, with comparisons against the DNS data of \cite{moser1999direct}. The solver was subsequently applied to density-stratified turbulent flows in complex geometries by \cite{rapaka2016immersed}. We further validate our numerical algorithm and the solver by comparing the $Nu$ from our simulations with those of \cite{pirozzoli2021towards} and \cite{de2022strong} for $Ra\leq 8\times 10^4$. Table \ref{tab:case details} presents the corresponding $Nu(Ra)$ values for the above-mentioned cases and are in very good agreement with the literature values. \\

\begin{table}
\centering
\rotatebox{90}{
\begin{minipage}{\textheight}
\centering
  \begin{tabular}{lcccccccc}
     $Case$& $Ra$ & $L_x\times L_y\times L_z$ & $N_{x}\times N_{y}\times N_{z}$ & $dz_{min}$ & $\delta_{\theta} $&$Nu$ & $Nu$ \citep{de2022strong} \\
     $1$ & $1\times 10^{3}$ & $1\times1\times1$ &  $384\times384\times256$ & $1.654\times 10^{-4}$ &8.52$\times 10^{-2}$&11.66& 11.14 \\
     $2$ &$2.5\times10^{3}$ & $1\times1\times1$ & $384\times384\times256$ &$1.654\times 10^{-4}$ &3.86 $\times 10^{-2}$& 28.97&28.56 \\
     $3$ &$5\times10^{3}$ &  $1\times1\times1$ &$384\times384\times256$ & $6.076\times 10^{-5}$ &2.06 $\times 10^{-2}$ & 53.08&52.20\\
     $4$ &$7.5\times10^{3}$ & $1\times1\times1$ &$384\times384\times256$& $6.076\times 10^{-5}$&1.06 $\times 10^{-2}$ & 77.85&75.73\\
     $5$ &$1\times10^{4}$ & $0.5\times0.5\times1$ & $384\times384\times256$ &$3.337\times10^{-5}$ & 8.85$\times 10^{-3}$&97.81&99.84\\
     $6$ &$3\times10^{4}$ &$0.3\times0.3\times1$&$384\times384\times256$&$1.279\times10^{-5}$ & 5.28 $\times 10^{-3}$& 296.13&281.14\\
     $7$ &$4\times10^{4}$ &$0.3\times0.3\times1$&$384\times384\times256$ & $1.279\times10^{-5}$& 1.68 $\times 10^{-3}$ & 403.03&370.17\\
     $8$ &$8\times10^{4}$ & $0.125\times0.125\times1$ & $384\times384\times256$ &$6.029\times10^{-6}$ & 1.01  $\times 10^{-3}$& 737.08&709.00\\
     $9$ & 1 $\times 10^{5}$  & $0.0625\times 0.0625\times1$ &  $384\times384\times768$ &$3.817\times10^{-6}$  &8.88 $\times 10^{-4} $&822.11&-\\
     $10$ &$2\times10^{5}$  & $0.0625\times 0.0625\times1$ &  $384\times384\times768$ &$3.817\times10^{-6}$ & 3.68 $\times 10^{-4}$& 1818.29&-\\
     $11$ &$2.5\times10^{5}$  & $0.0625\times 0.0625\times1$ & 
     $384\times384\times768$ & $3.817\times 10^{-6}$& 2.23 $\times10^{-4}$ &2264.8&-\\
     $12$ &$4\times10^{5}$  & $0.015625\times 0.015625\times1$ &  $384\times384\times768$ &8.354 $\times 10^{-7}$&$2.54\times 10^{-4}$& 2585.61 &-\\
     $13$ &$5\times10^{5}$  & $0.015625\times 0.015625\times1$ &  $384\times384\times768$ &8.354 $\times 10^{-7}$ &1.98 $\times 10^{-4}$ & 3188.14 &-\\
     $14$ &$6\times10^{5}$ & $0.015625\times 0.015625\times1$ & $384\times384\times768$ & 8.354 $\times 10^{-7}$& 1.59$\times10^-4 $&3920.78&- \\
     $15$ &$7\times10^{5}$ &$0.015625\times 0.015625\times1$&$384\times384\times768$& 8.354 $\times 10^{-7}$ & 1.35 $\times10^{-4}$& 4702.81 &-\\
     $16$ &$8\times10^{5}$  &$0.015625\times 0.015625\times1$&$384\times384\times768$ &8.354 $\times 10^{-7}$& 1.06 $\times 10^{-4}$& 5645.41 &-\\
     $17$ &$9\times10^{5}$  &$0.015625\times 0.015625\times1$&$384\times384\times768$ &8.354 $\times 10^{-7}$&9.57$\times10^{-5}$ & 6204.75 &-\\
     $18$ &$1\times10^{6}$  &$0.015625\times 0.015625\times1$&$384\times384\times768$ &1.22 $\times 10^{-7}$& 7.97$\times 10^{-5}$& 7514.61&-\\
   
     

\end{tabular}
\caption{The cases correspond to different Rayleigh numbers ($Ra$), their corresponding domain lengths (variable in horizontal directions ($L_x$ and $L_y$) while fixed in vertical ($L_z=1$)), grid details, smallest grid spacing in vertical direction ($dz_{min}$), the thermal boundary layer thickness ($\delta_{\theta}$), the Nusselt number($Nu$) and the $Nu$ from \cite{de2022strong}.}
\label{tab:case details}
\end{minipage}
}
\end{table}

\section{Results} \label{sec:section3}

We investigate the heat transfer behavior and its scaling characteristics in section \ref{sec:section3.1} and discuss the flow dynamics in section \ref{sec:section3.2} of the Rayleigh-Darcy convection in fluid-saturated three-dimensional porous media for $1\times10^3 \leq Ra \leq 1\times 10^6$. The details of the cases are listed in Table \ref{tab:case details}. The boundary layer thickness, computed based on the variance of the temperature, remains significantly smaller than the size of the domain in the horizontal direction, justifying the choice of $L_x$ and $L_y$ for all the cases \citep{iyer2020classical}. We further use the temperature statistics in section \ref{sec:section3.3} to identify the transition to the ultimate regime, and the corresponding scaling behavior. In section \ref{sec:section3.4}, we analyze the morphology of the near-boundary and mid-plane to characterize the dominant flow structures. The characteristic length scale was determined using the energy distribution across horizontal modes and its relation to $Ra$. \\

\subsection{Heat transfer and Scaling behavior} \label{sec:section3.1}
In natural convection, the relationship between the Nusselt number $(Nu)$ and the Rayleigh number $(Ra)$ is described by a power law $Nu \sim Ra^{\alpha}$, where the exponent $\alpha$ characterises the structure of the convective flow. This exponent in Rayleigh-B\'enard convection (RBC) varies from 1/3 to 1/2, indicating a transition from moderate to high $Ra$ values corresponding to the ultimate regime \citep{kraichnan1962turbulent}. However, for the Rayleigh-Darcy convection, the $Nu$ scales linearly with $Ra$ from the intermediate
range \cite{otero2004high} to the ultimate regime \citep{pirozzoli2021towards}. The reason for this linear scaling is attributed to the confinement of the temperature gradients within a thickness $\delta{\color{black}_{\theta}} \propto Ra^{-1}$ near the solid boundary and well-mixed interior of the domain (negligible thermal gradient) at high Ra. Since $Nu$ is computed from the temperature gradients, it is inversely proportional to $\delta{\color{black}_{\theta}}$, and hence $Nu \propto Ra$. \\

\begin{figure}
    \centering
    \begin{subfigure}[b]{\textwidth}
         \includegraphics[width=\linewidth]{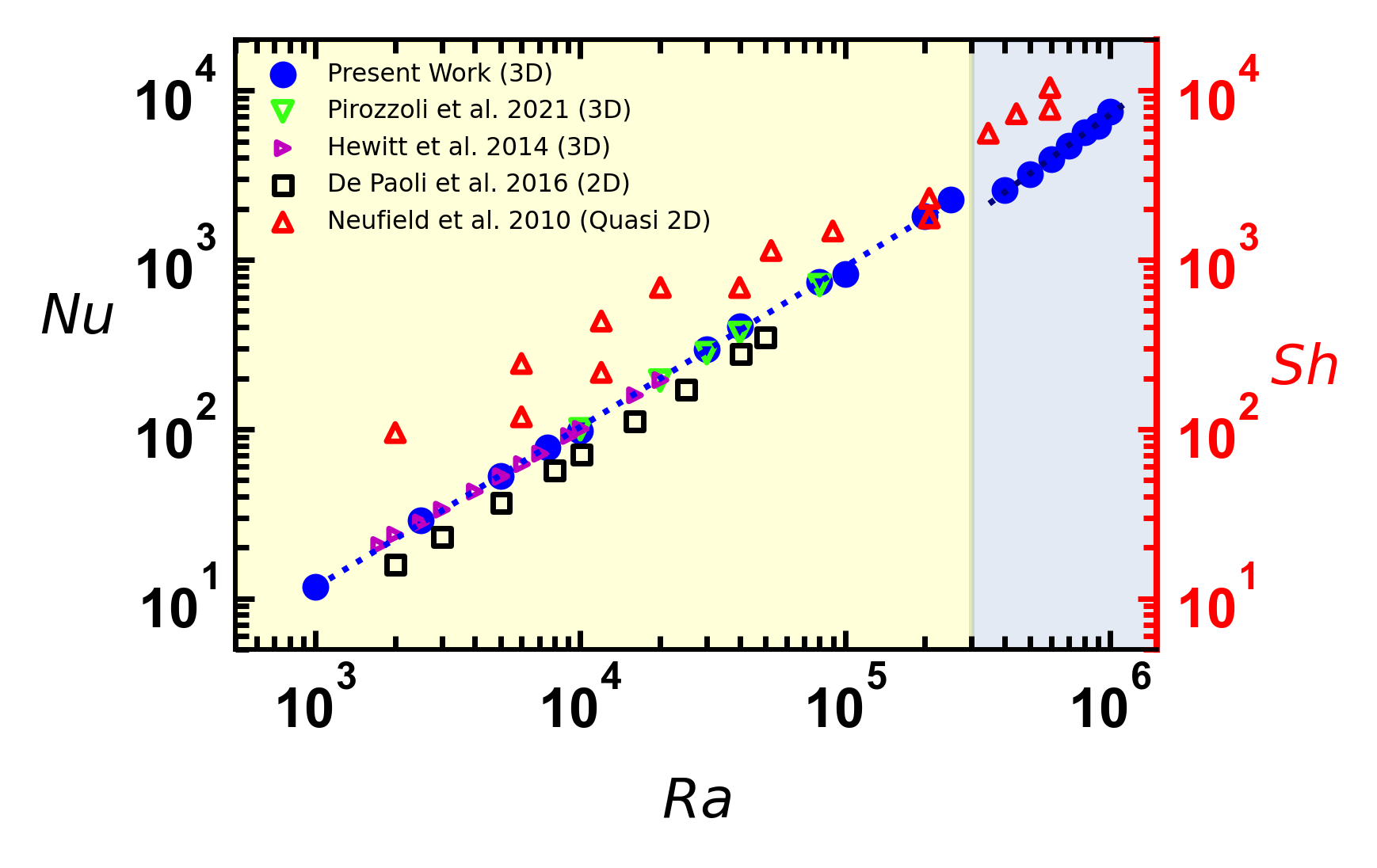}
        \caption{\label{fig:2a}}
    \end{subfigure}
    \hfill
    \begin{subfigure}{\linewidth}
         \includegraphics[width=\linewidth]{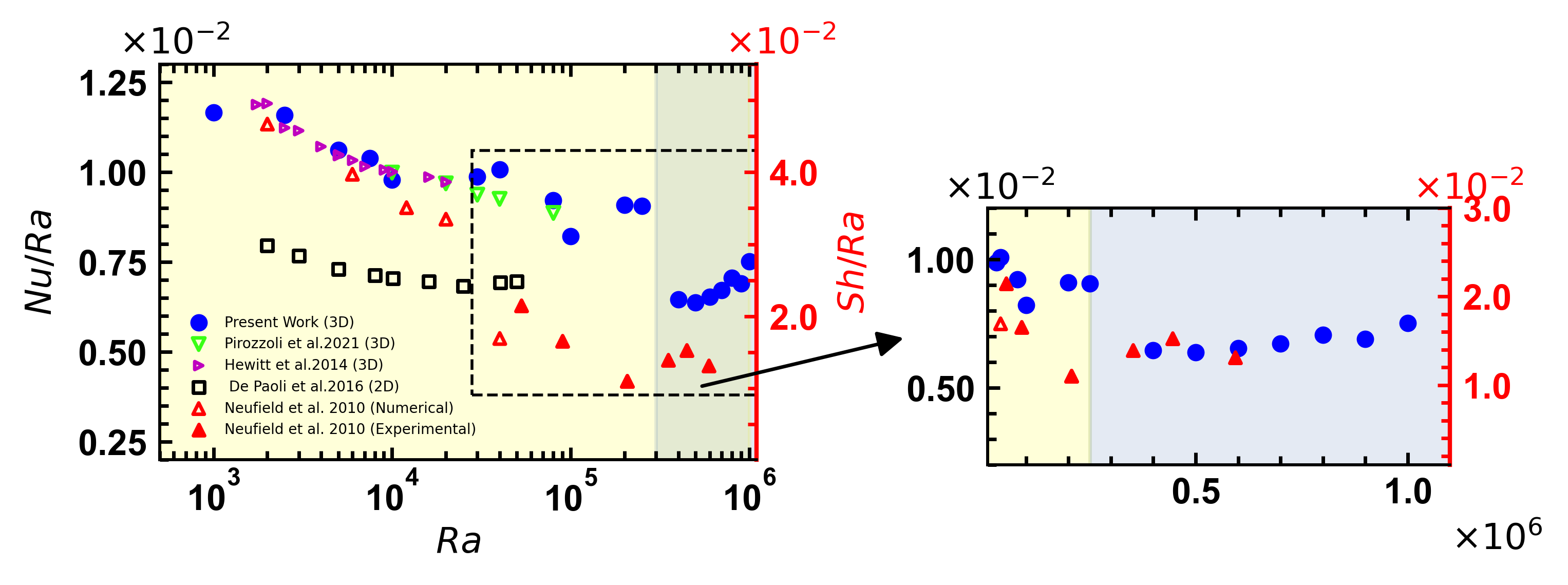}
        \caption{\label{fig:2b}}
    \end{subfigure}
     \captionsetup{justification=justified, singlelinecheck=false,width=\textwidth}
    \caption{\label{fig:2} (a) The variation of the time-averaged Nusselt number ($Nu$) in relation to the Rayleigh number ($Ra$) for the current numerical simulation (shown by filled blue circles) within the range of $1\times10^3\leq Ra\leq 1\times10^6$. The best fit is $Nu=0.009Ra+8.27$ ($R^2$=0.9987) for $Ra\leq2.5\times10^5$ and $Nu$ = 0.008$Ra$-805.94 ($R^2$=0.9898) for $Ra\geq 4\times10^5$. Results obtained in previous studies are shown by different open symbols (\cite{pirozzoli2021towards} inverted green triangle),(\cite{hewitt2014high} rightward triangle), (\cite{de2016influence} black square). 
    Red triangles \citep{neufeld2010convective} represent the variation of the Sherwood number ($Sh$) (right ordinate) with respect to $Ra$. (b) Variation of time-averaged Nusselt number ($Nu$) scaled by $Ra$ (termed as compensated Rayleigh number ($Nu/Ra$), with respect to $Ra$. 
    }
\end{figure}

We quantify the heat flux in terms of the horizontally and time-averaged Nusselt number ($Nu$; equation \ref{Eq:Nu}) for all cases, using the temperature gradient at the boundaries.  The time variation of the Nusselt number exhibits chaotic fluctuations about the time-averaged value $Nu$. For all the cases, time averaging is performed over a period when a statistically steady state is achieved. The Nusselt numbers obtained from our simulations are shown in Table \ref{tab:case details}. We plot these $Nu$ (the filled blue circles) values with respect to $Ra$ in figure \ref{fig:2a}. The values of $Nu$ from the literature (different open symbols) are also plotted for comparison. The  $Nu$ values of our simulations are in good agreement with the existing 3D numerical results of \cite{pirozzoli2021towards} up to $8\times10^4$, \cite{hewitt2014high} up to $2\times10^4$. However, our $Nu$ values are higher than those in the 2D simulations \citep{de2016influence} for the given $Ra$ value. The right ordinate, presented in red colour, represents the convective dissolution of carbon dioxide in saline aquifers in terms of the Sherwood number $(Sh)$, which is analogous to thermal dissolution and is used to compare the results of \cite{neufeld2010convective}.\\

We find a linear scaling for the $Nu(Ra)$ dataset spanning the range $1\times 10^3 \leq Ra \leq 1\times 10^6$. A transition in the linear scaling could also be observed from $Ra\sim 4 \times 10^5$. We use different colours to indicate this change in the scaling in Figure \ref{fig:2a}. This transition is the upshot of the dominance of buoyancy forces over dissipative forces, leading to boundary layer instabilities \citep{graham1994plume}. These instabilities contribute to the formation of proto-plumes within the domain, ultimately disrupting the boundary layer and resulting in the emergence of unstable convective rolls. This transition in scaling indicates the beginning of the ultimate regime in Rayleigh-Darcy convection. We linear fit our dataset as shown in equations \ref{eq:nu_scaling}\hyperlink{partE}{a} and \ref{eq:nu_scaling}\hyperlink{partE}{b}.\\
\begin{subequations} \label{eq:nu_scaling}
\begin{align}
{{Nu}} = {\alpha{Ra}}+{\beta};\quad
\alpha = 0.009,\quad \beta=8.27.\\
{{Nu}} = {\gamma{Ra}}+{\zeta};\quad
\gamma = 0.008,\quad \zeta=-805.94
\end{align}
\end{subequations}

{\color{black}The scaling exponent for $Ra \leq 2.5 \times 10^5$ (equation \ref{eq:nu_scaling}\hyperlink{partE}{a}) {\color{black} is $\sim12.5\%$ higher than for the scaling in the ultimate regime (equation \ref{eq:nu_scaling}\hyperlink{partE}{b}), $Ra > 2.5 \times 10^5$. Our scaling is $\sim 6.25\%$ lesser than the scaling of \cite{hewitt2014high} ($Nu=0.0096Ra+4.6$ given for $Ra \leq 2\times 10^4$). Moreover, our results are $\sim1.24 \%$ lower than the extrapolated prediction of \cite{pirozzoli2021towards} for the ultimate regime ($Nu=0.0081Ra$ for $Ra\geq 5\times 10^5$), demonstrating a close agreement.
}\\}

To further identify the transition to the ultimate regime, we scaled the Nusselt number ($Nu$) with the Rayleigh number ($Ra$) as \enquote*{$Nu/Ra$} (compensated $Nu$) and plotted it as a function of \enquote*{$Ra$} as shown in figure \ref{fig:2b}. Similarly, $Sh$ values from \cite{neufeld2010convective} are also scaled with $Ra$ and plotted using the right ordinate with red-coloured triangles for their quasi-2D analysis. The current simulation results (filled blue circles) agree very well with the results of \cite{pirozzoli2021towards, hewitt2014high} for $ Ra \leq 8 \times 10^4$. The difference in the linear scaling before and after $Ra 
\sim
4
\times 10^5$ suggests a transition to the ultimate regime, characterized by approximately constant values of the compensated Nusselt number $Nu/Ra$. A similar transition in the compensated Sherwood number could also be observed $\sim Ra >2\times10^4$ from the quasi-2D numerical ($Ra \leq 4 \times10^4$) and experimental ($Ra \sim 10^6 $) datasets of \cite{neufeld2010convective}. \\

This plateau of the compensated $Nu$ suggests the onset of a fully developed convective state, where most of the temperature (or concentration) drop is confined within thin boundary layers, while the interior remains nearly well-mixed. This behavior is consistent with the prediction of linear scaling $Nu \sim Ra$ \citep{otero2004high} or $Sh \sim Ra$ \citep{liyanage2019multidimensional}, which is the characteristic of the ultimate regime of porous convection. This proposed linear scaling $Nu \sim Ra$ is a deviation from the conventional $1/3$ scale observed in RBC. This difference arises owing to the behavior of the thermal layer thickness. In RBC, the thermal boundary layer thickness contributes to the Rayleigh number with an exponent of $3$. In contrast, for RDC, the thermal boundary layer thickness appears with an exponent of $1$, along with the remaining two powers of the length scale, represented by the squares of the pore length scale through Darcy permeability ($\kappa$)(Equation \ref{Eq:Ra}).\\

\begin{figure}
    \centering
    \includegraphics[width=0.7\linewidth]{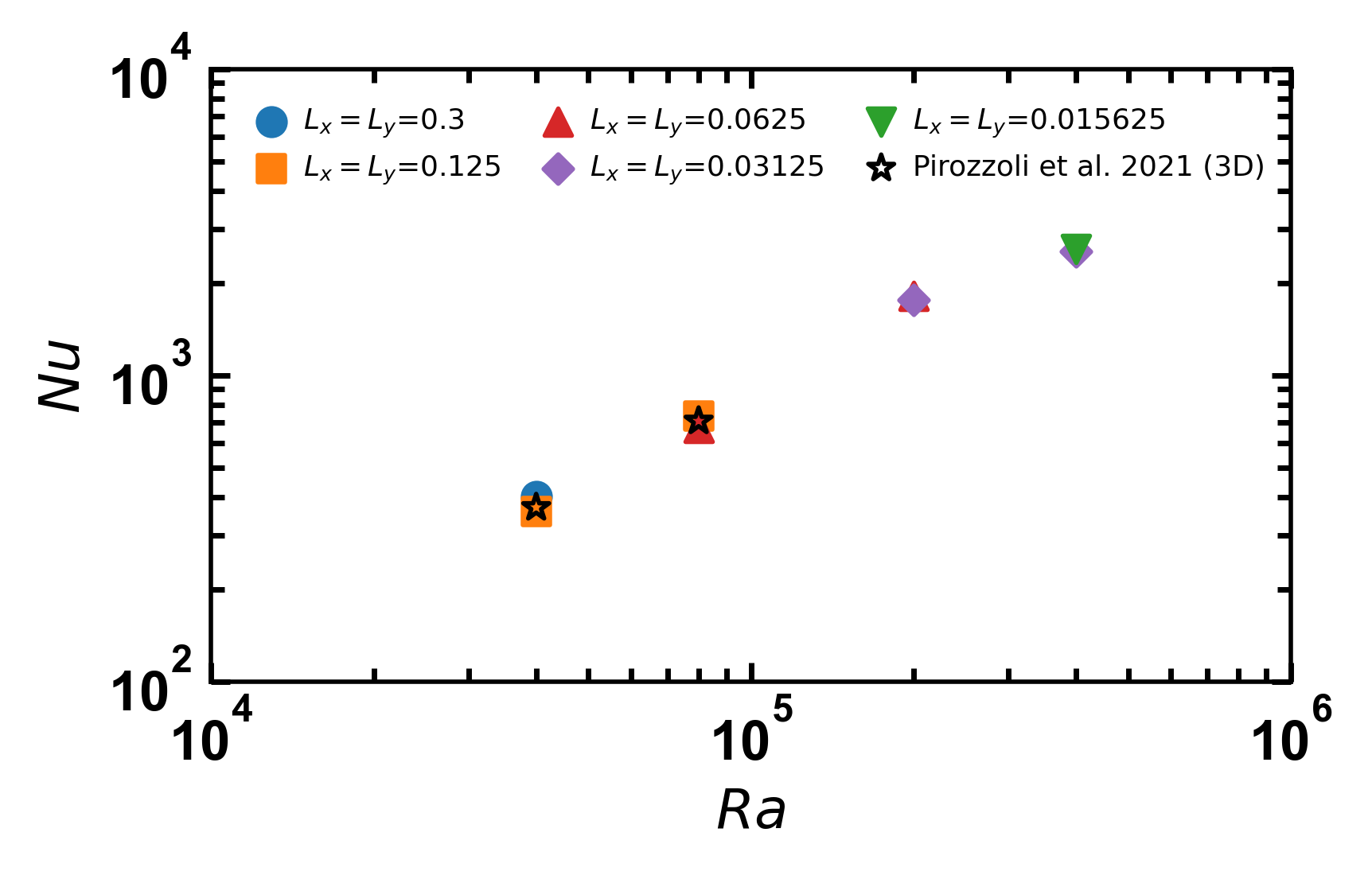}
    \caption{The time-averaged Nusselt number ($Nu$) is presented for various $Ra$ values across different domain sizes, along with a comparison to the results reported in \cite{pirozzoli2021towards}.}
    \label{fig:domain_independency}
\end{figure}

To assess the influence of domain size in the horizontal direction on $Nu$, we vary $L_x=L_y$ while keeping the vertical height constant for a particular $Ra$ value. Figure \ref{fig:domain_independency} compares the time-averaged $Nu$ for different horizontal domain sizes. The $Nu$ values at $Ra = 4 \times 10^4 $ and $ 8\times 10^4$ from \cite{pirozzoli2021towards} are also included as reference. The variation in $Nu$ across different horizontal domain sizes at a given $Ra$ is small \citep{hewitt2014high} for all cases and compares well with the available Nu values reported by \cite{pirozzoli2021towards}. These results further corroborate our choice of lower-aspect-ratio domains for high-$Ra$ cases to utilize computational resources efficiently.\\

\subsection{Flow Dynamics} \label{sec:section3.2}

\begin{figure}
    \begin{subfigure}[b]{0.3\linewidth}
         \includegraphics[width=\linewidth]{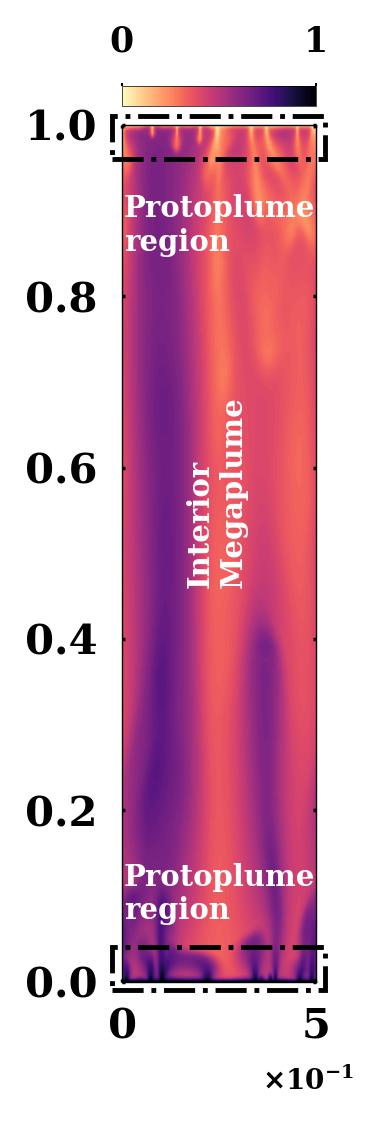}
          \caption{}
    \end{subfigure}
    \hfill
     \begin{subfigure}[b]{0.3\linewidth}
\includegraphics[width=1.05\linewidth]{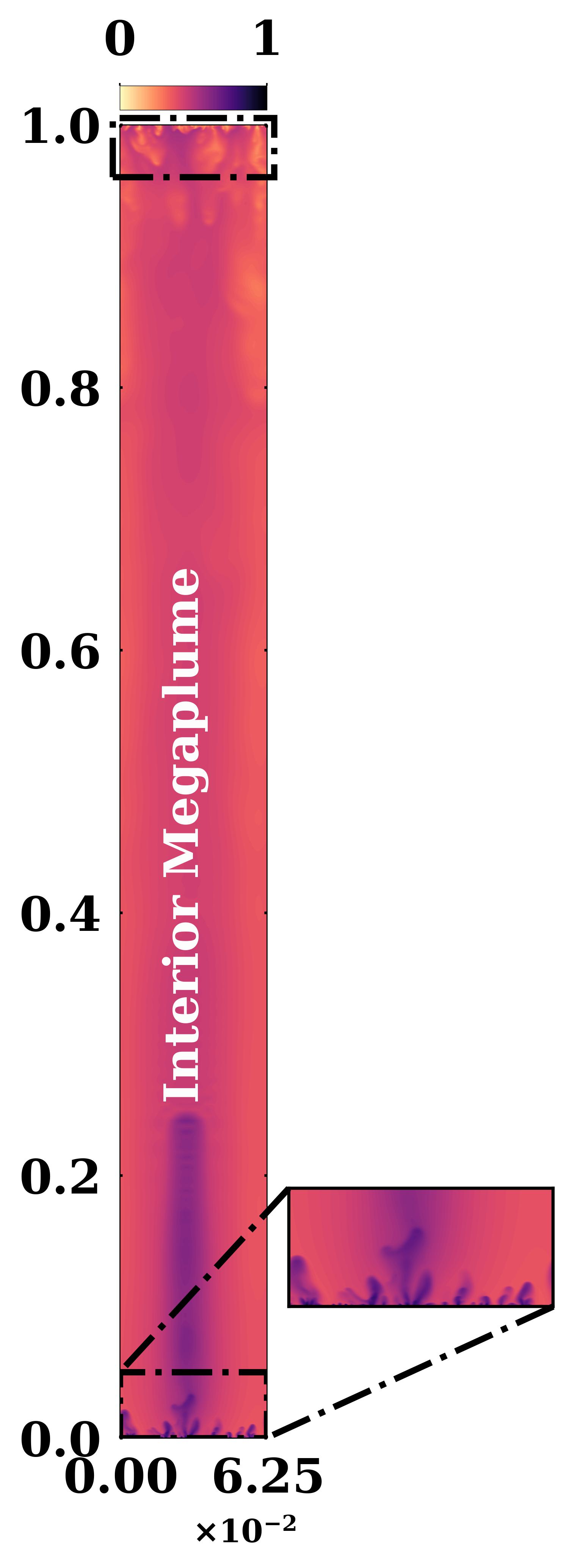}
 \caption{}
    \end{subfigure}
    \hfill
     \begin{subfigure}{0.3\linewidth}
\includegraphics[width=0.8\linewidth]{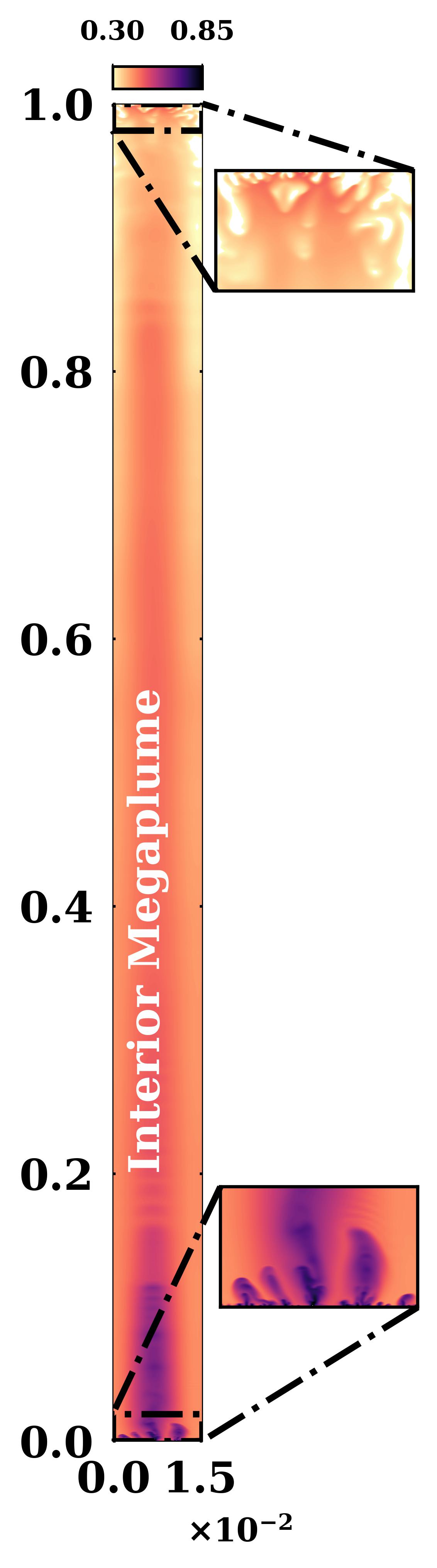}
 \caption{}
    \end{subfigure}
 \captionsetup{justification=justified, singlelinecheck=false,width=\textwidth}
    \caption{\label{midplane} The temperature distribution at the midplane of the horizontal direction shows the protoplume region near the boundary (captured by black dotted boxes near the upper and lower boundary) and extensive megaplumes across the domain for (a)$Ra$=$1\times10^4$ (b) $Ra$=$2.5 \times 10^5$ (c)$1\times 10^6$.}
\end{figure}

\begin{figure}  
\begin{subfigure}[b]{0.49\linewidth}
     \includegraphics[width=\linewidth, height=0.3\textheight]{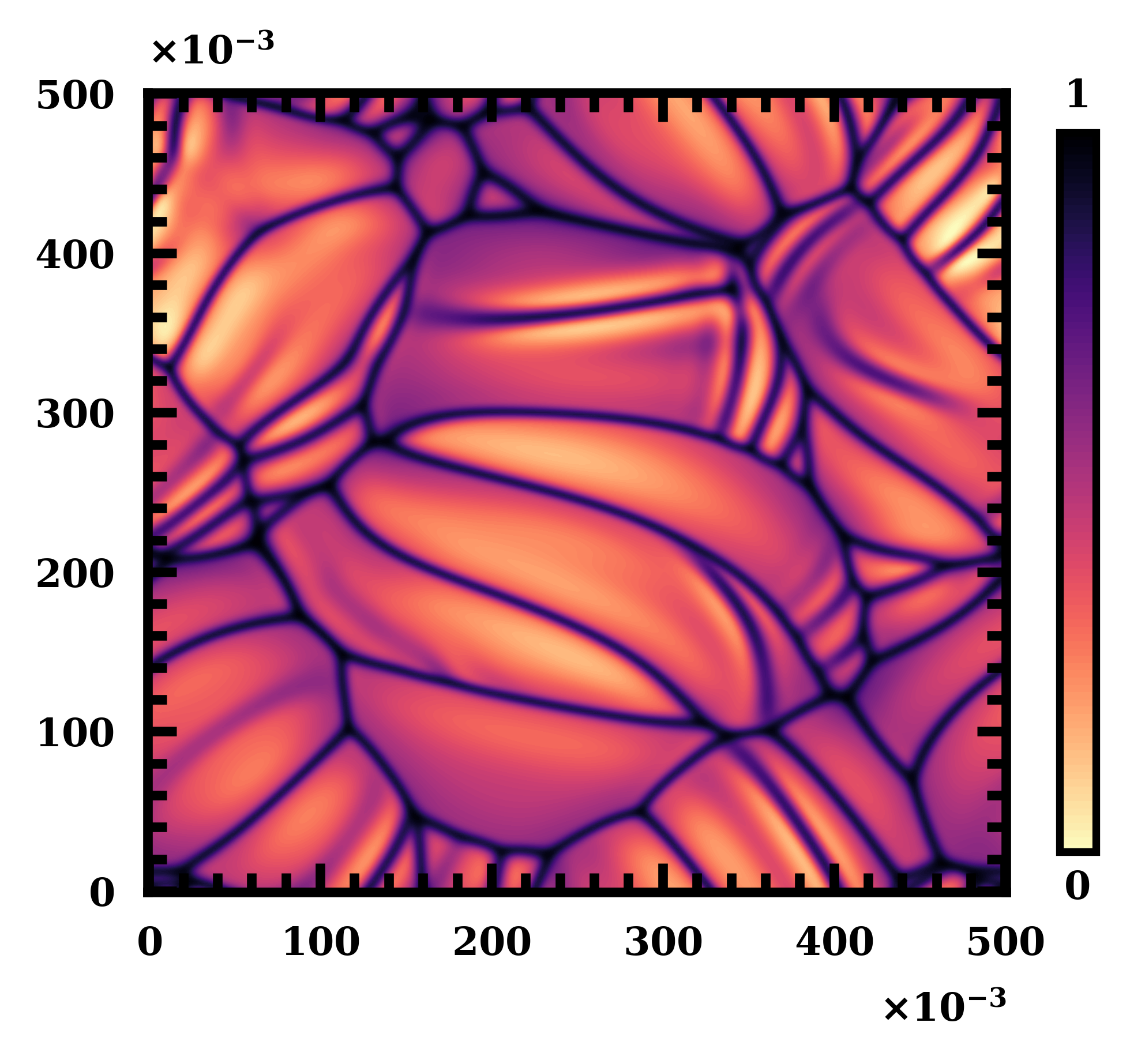}
    \caption{}
\end{subfigure} 
\begin{subfigure}[b]{0.49\linewidth}
     \includegraphics[width=\linewidth, height=0.3\textheight]{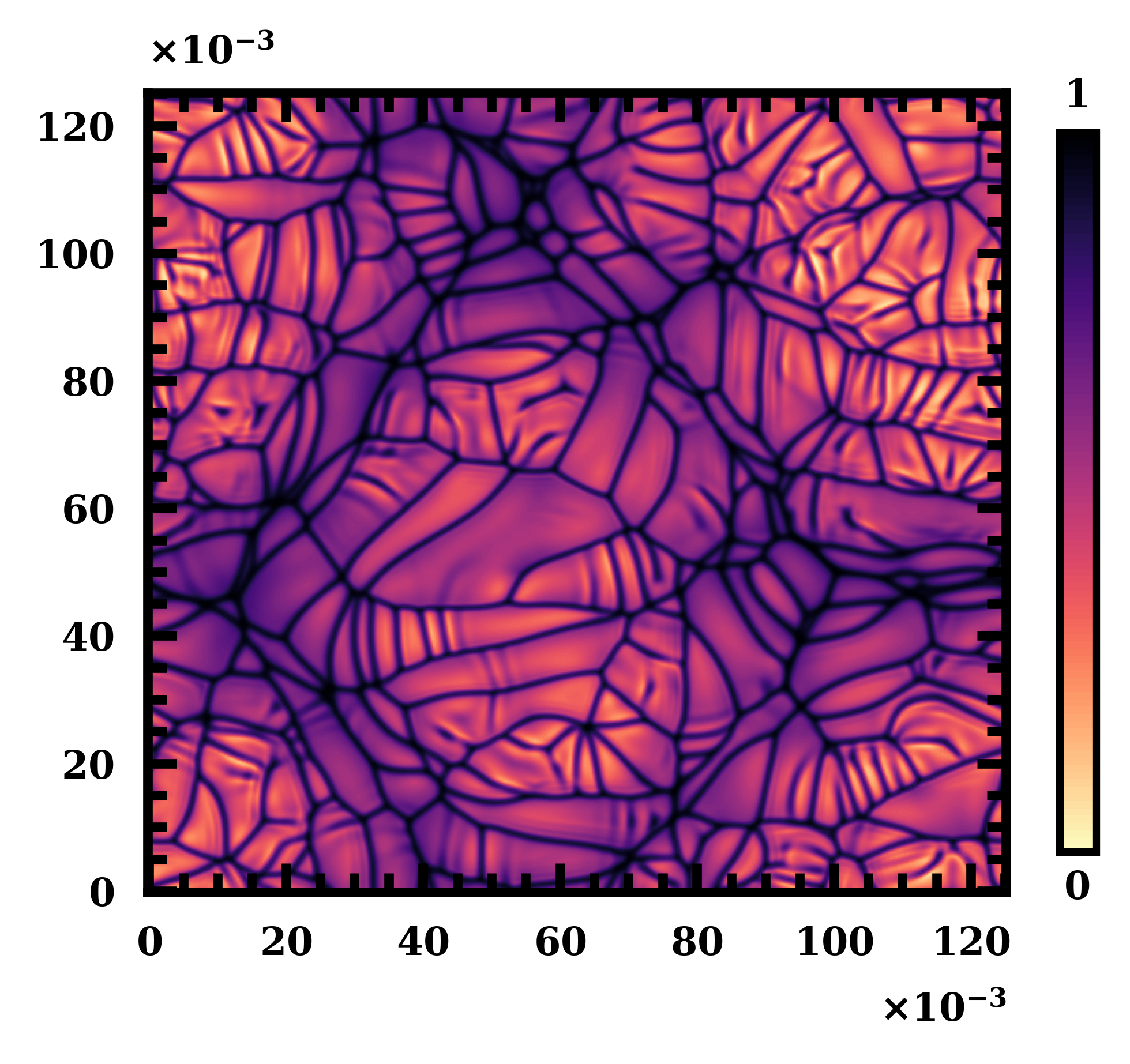}
    \caption{}
\end{subfigure}
\begin{subfigure}[b]{0.49\linewidth}
     \includegraphics[width=\linewidth, height=0.3\textheight]{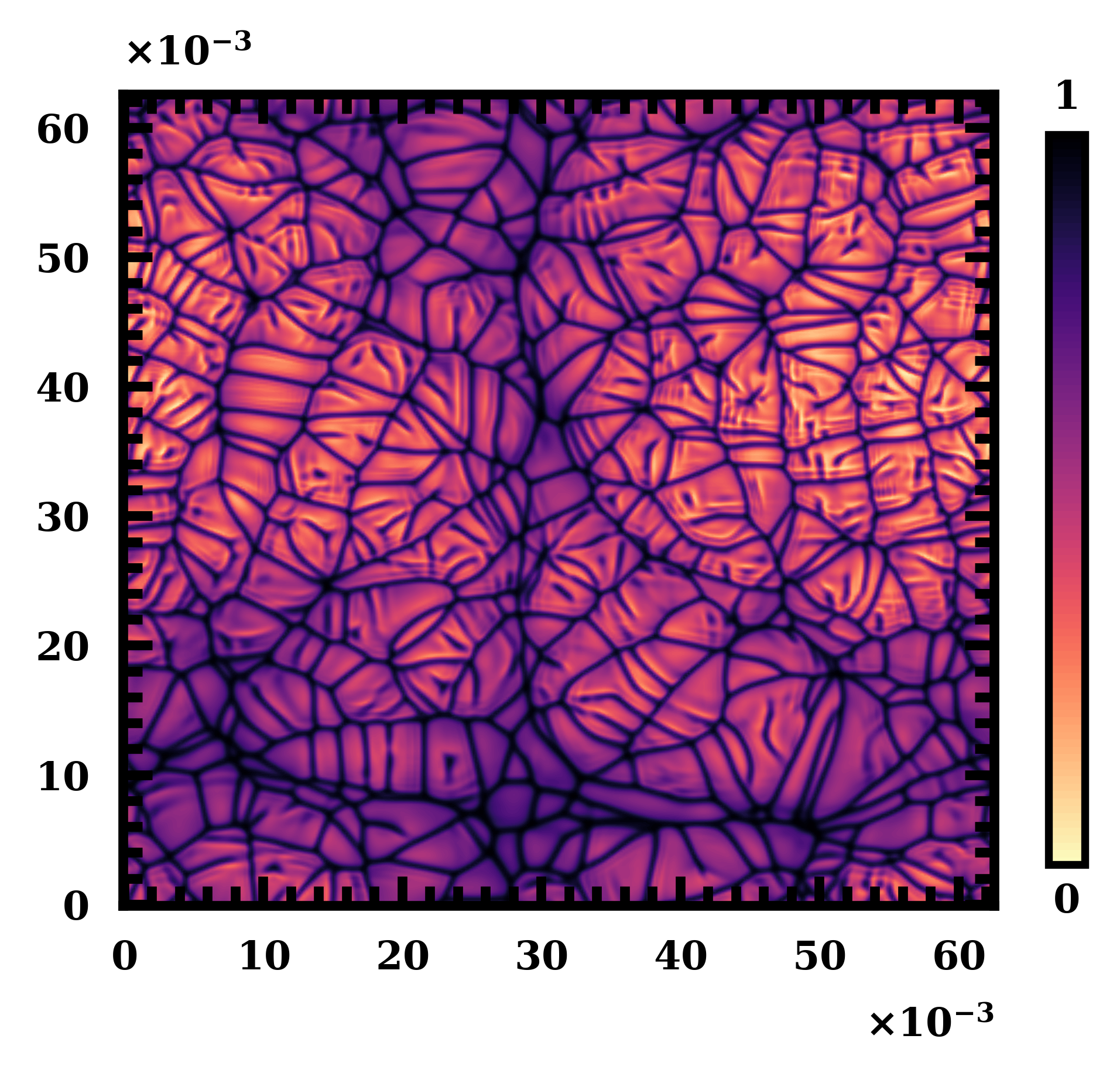}
    \caption{}
\end{subfigure}
\begin{subfigure}[b]{0.49\linewidth}
     \includegraphics[width=\linewidth, height=0.3\textheight]{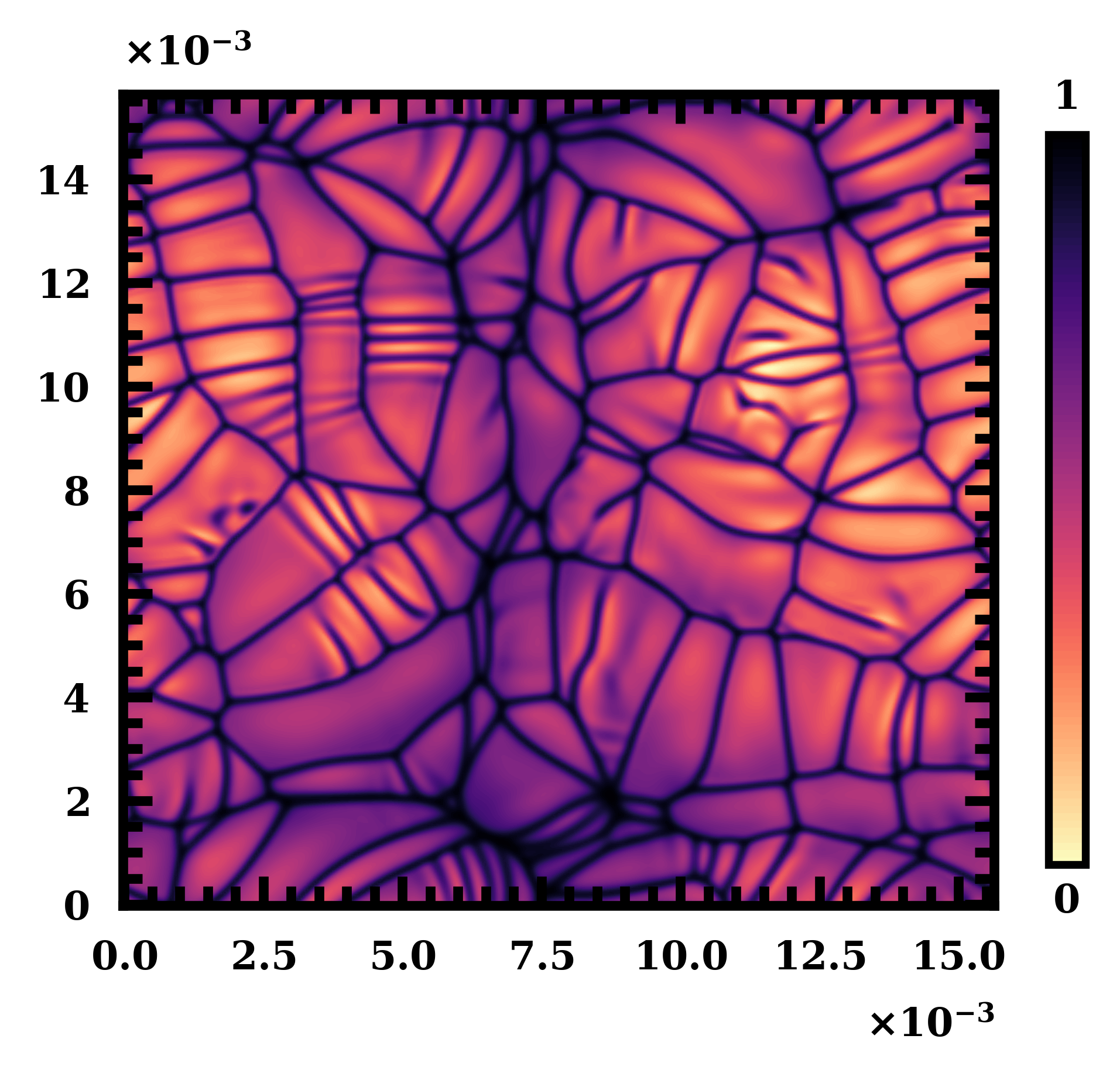}
    \caption{}
\end{subfigure}
 \begin{subfigure}[b]{0.49\linewidth}
     \includegraphics[width=\linewidth, height=0.3\textheight]{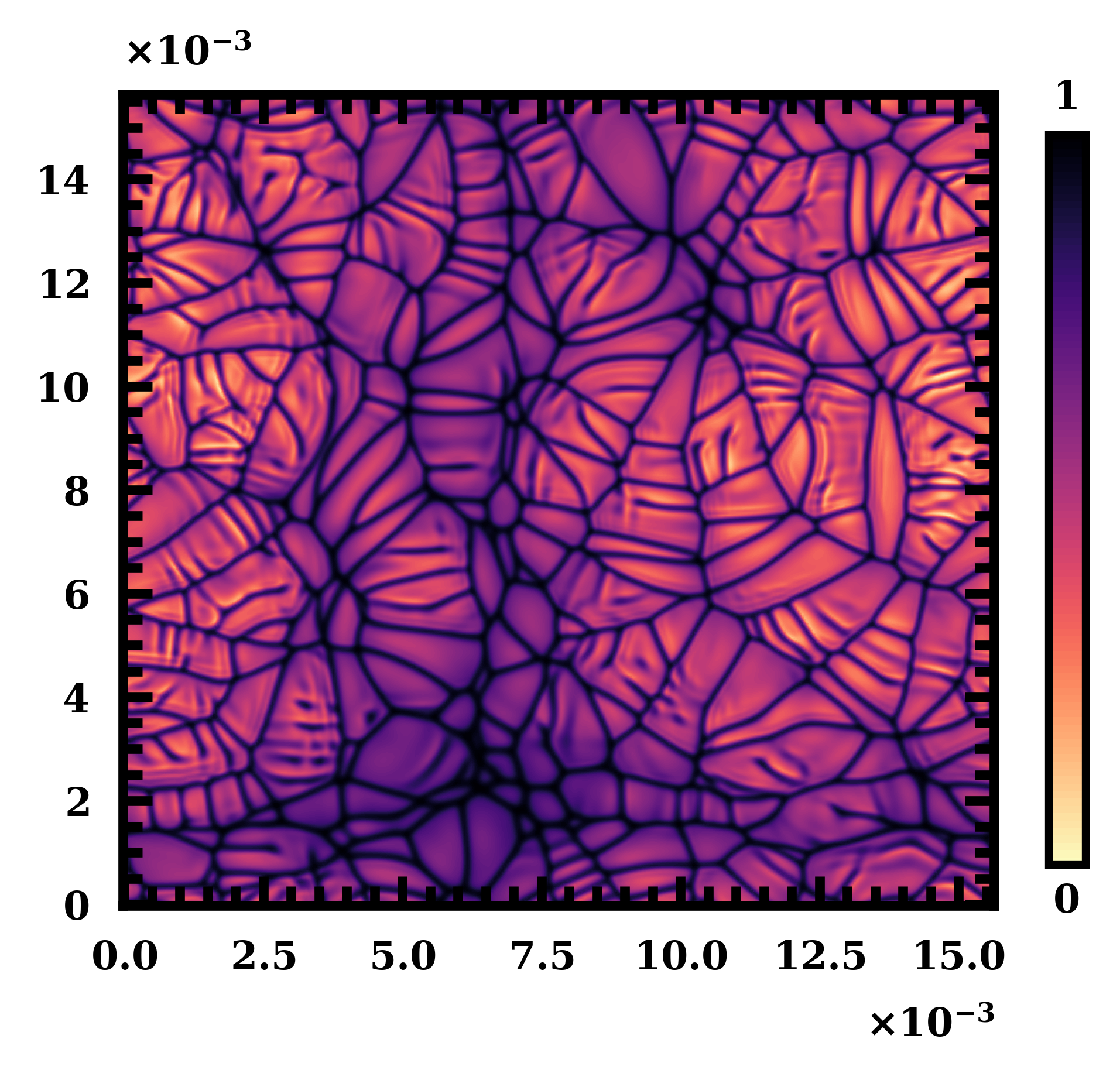}
    \caption{}
\end{subfigure} 
\begin{subfigure}[b]{0.49\linewidth}
\includegraphics[width=\linewidth,height=0.3\textheight]{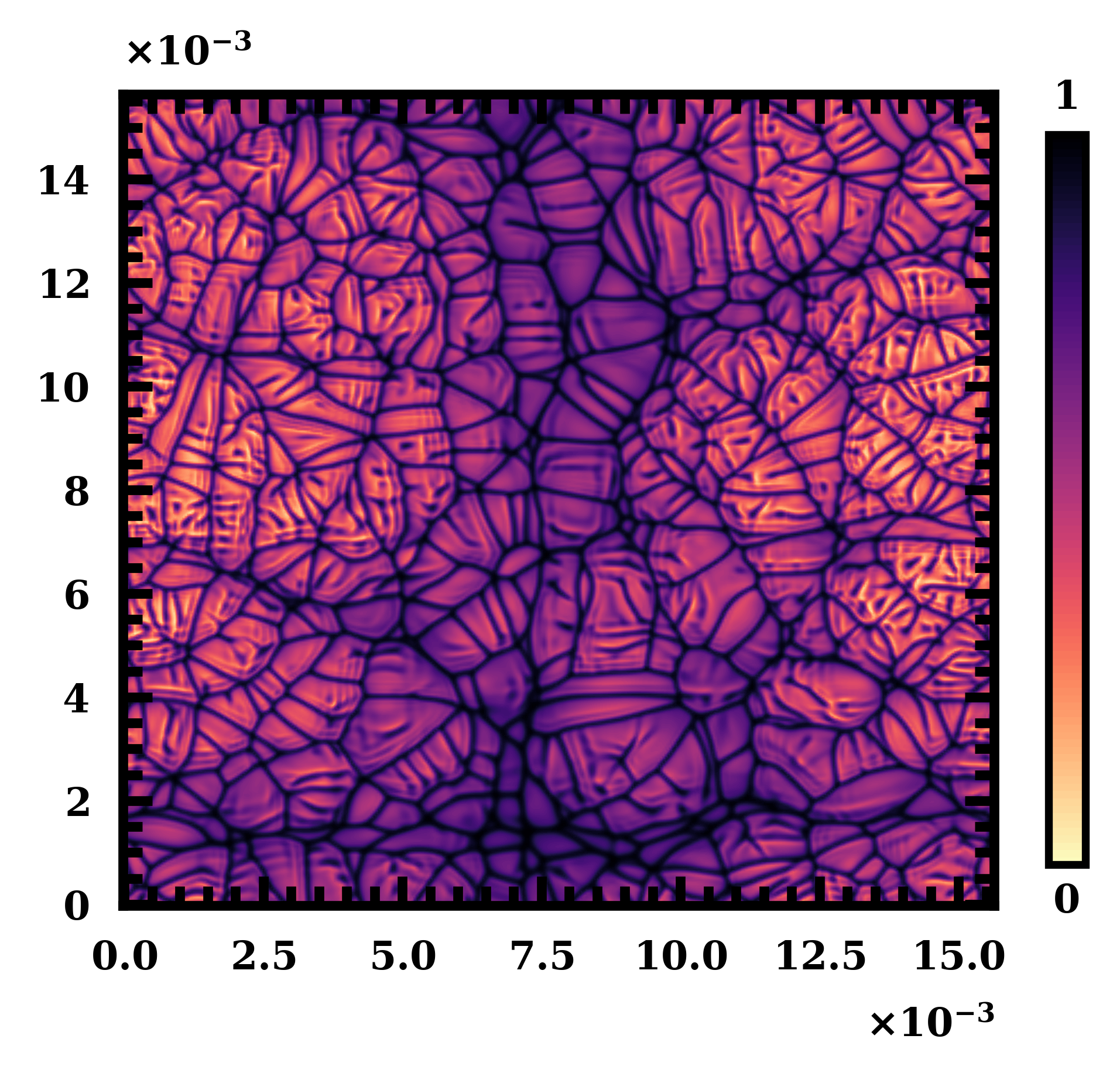}
    \caption{}
\end{subfigure}
 \captionsetup{justification=justified, singlelinecheck=false,width=\textwidth}

\caption{\label{fig:4} Instantaneous dimensionless temperature distribution at the near-wall plane ($z=30/Ra$) for \textit{(a)} $Ra= 1\times10^4$, \textit{(b)} $Ra= 8\times10^4$, \textit{(c)} $Ra= 2\times 10^5$, \textit{(d)} $Ra= 6\times 10^5$, \textit{(d)} $Ra= 8\times 10^5$, and \textit{(e)} $Ra= 1\times 10^6$. }

\end{figure}\hypertarget{partE}{}

The change in $Nu$ with $Ra$ implies the corresponding change in the flow structure—ranging from organized convective rolls to chaotic, boundary-driven plume shedding. This behavior arises from the relative strength of advection and dissipation effects; therefore, at low $Ra$, viscous and diffusive forces dominate, suppressing convective motion and preventing plume formation \citep{de2022strong}. As $Ra$ increases,  buoyancy-driven advection becomes significant enough to initiate and sustain convective plumes. At high Rayleigh numbers, the flow exhibits unsteady dynamics characterized by vertically oriented, large-scale exchange motions, referred to as \enquote*{megaplumes}. These arise from the lateral aggregation of smaller filamentary structures, or \enquote*{protoplumes}. Figures \ref{midplane} (a), (b), and (c) show the vertical midplane slice to visualize both megaplumes and protoplumes for $Ra = 1\times10^4$, $2.5 \times 10^5$, and $1\times10^6$, respectively. With an increase in $Ra$, the number of protoplumes increases, which indicates enhanced convection from the boundaries and efficient heat transport. These protoplumes eventually coalesce to form a vertically aligned megaplume.\\ 


We also plot the thermal signatures of the protoplumes and the megaplumes in Figure \ref{fig:4} for different cases at $z = 30/Ra$ to assess the flow dynamics near the bottom boundary. The protoplumes, represented by short, thin, and bright filament-like structures that are interconnected and arranged into cells of different polygonal shapes, carry warm fluid from the bottom boundary. Within these cells, the colder fluid, represented by the lighter color, flows in to replace the hot fluid (shown by the darker color). This colder fluid, upon impacting the bottom boundary, deflects horizontally, further leading to interactions among the protoplumes. This results in clustering in specific regions bounded by thick, dark ridges. These thick, dark ridges signify the boundaries of the megaplumes. \\

\subsection{Temperature statistics} \label{sec:section3.3}

\begin{figure}
     \begin{subfigure}{\linewidth}
        \includegraphics[width=\linewidth]{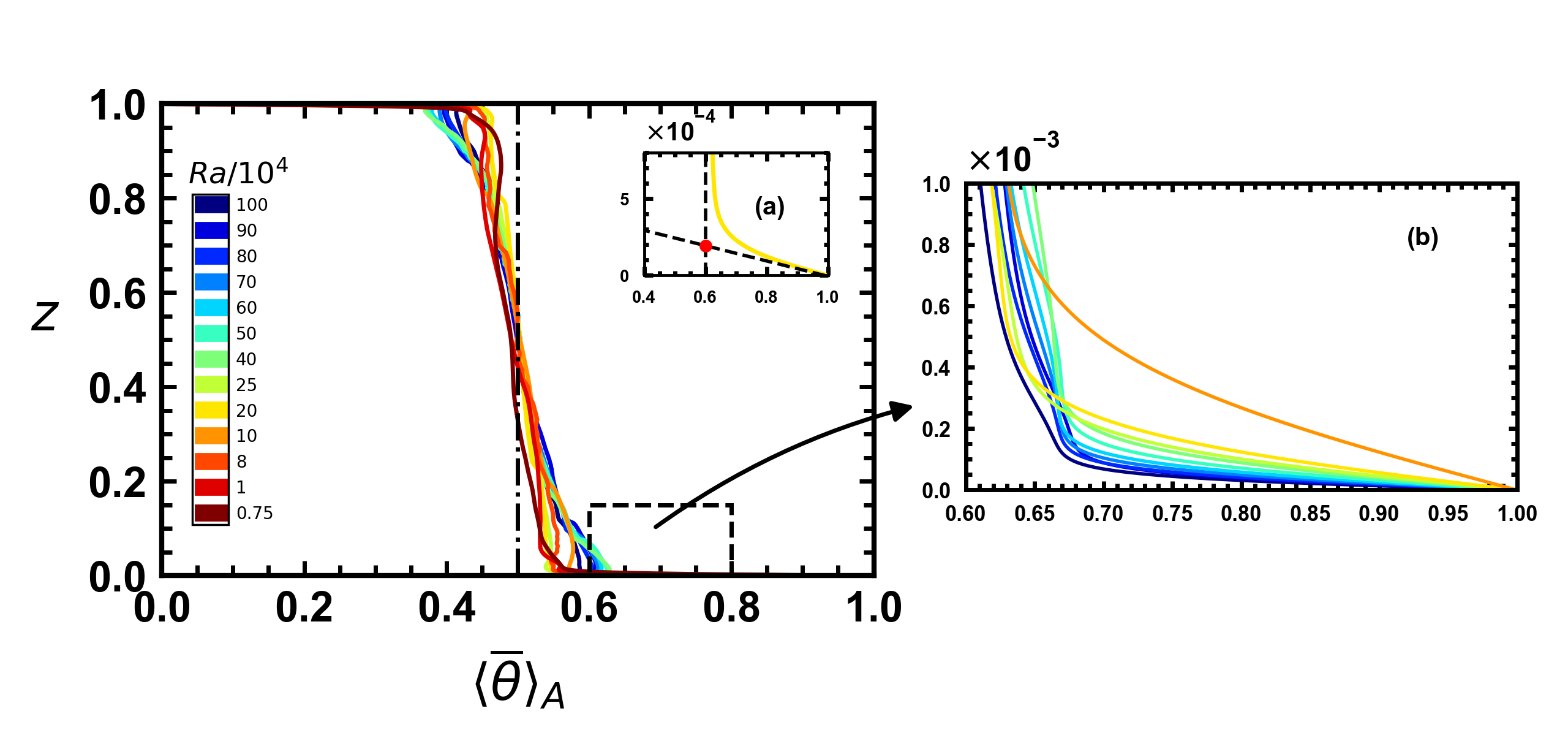}
       \end{subfigure}     
 \captionsetup{justification=justified, singlelinecheck=false,width=\textwidth}
    \caption{\label{fig:5} The time-averaged and horizontally averaged temperature $\langle \overline{\theta} \rangle$ distribution in the vertical direction  $(z)$ shows a relatively weak linear temperature gradient across the domain. Inset (a) shows the boundary region by the intersection of the linear profile fitting the bulk and near-wall regions; (b) zoomed view of the near-wall region showing the boundary layer thickness with respect to $Ra$ for $Ra \geq 1\times 10^5$.}
\end{figure}

\begin{figure}
   
     \begin{subfigure}{0.5\linewidth}
        \includegraphics[width=\linewidth]{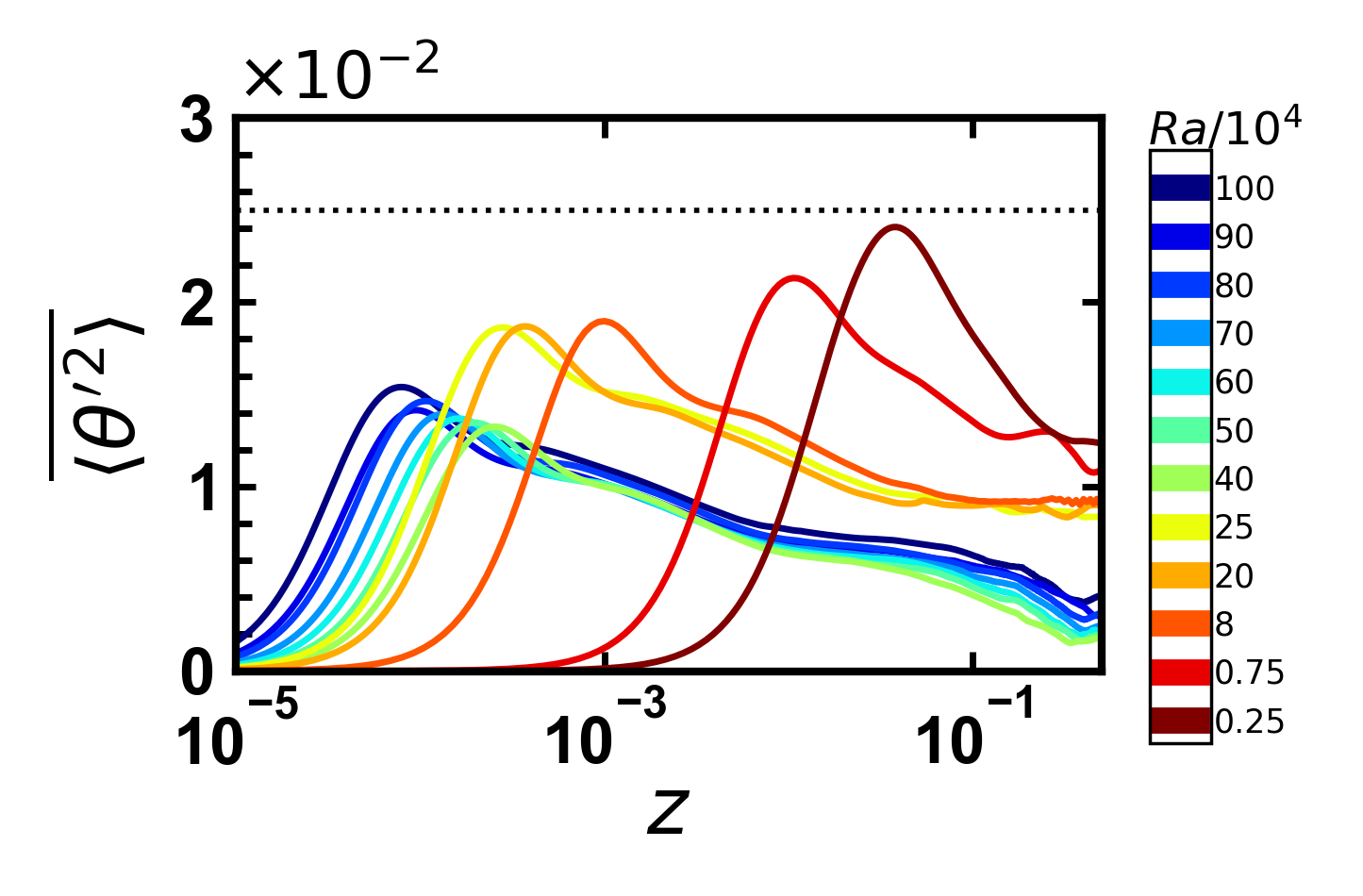}
        \caption{\label{fig:6a}}
    \end{subfigure}
 \captionsetup{justification=justified, singlelinecheck=false,width=\textwidth}
\hfill
 \begin{subfigure}{0.5\linewidth}
        \includegraphics[width=\linewidth]{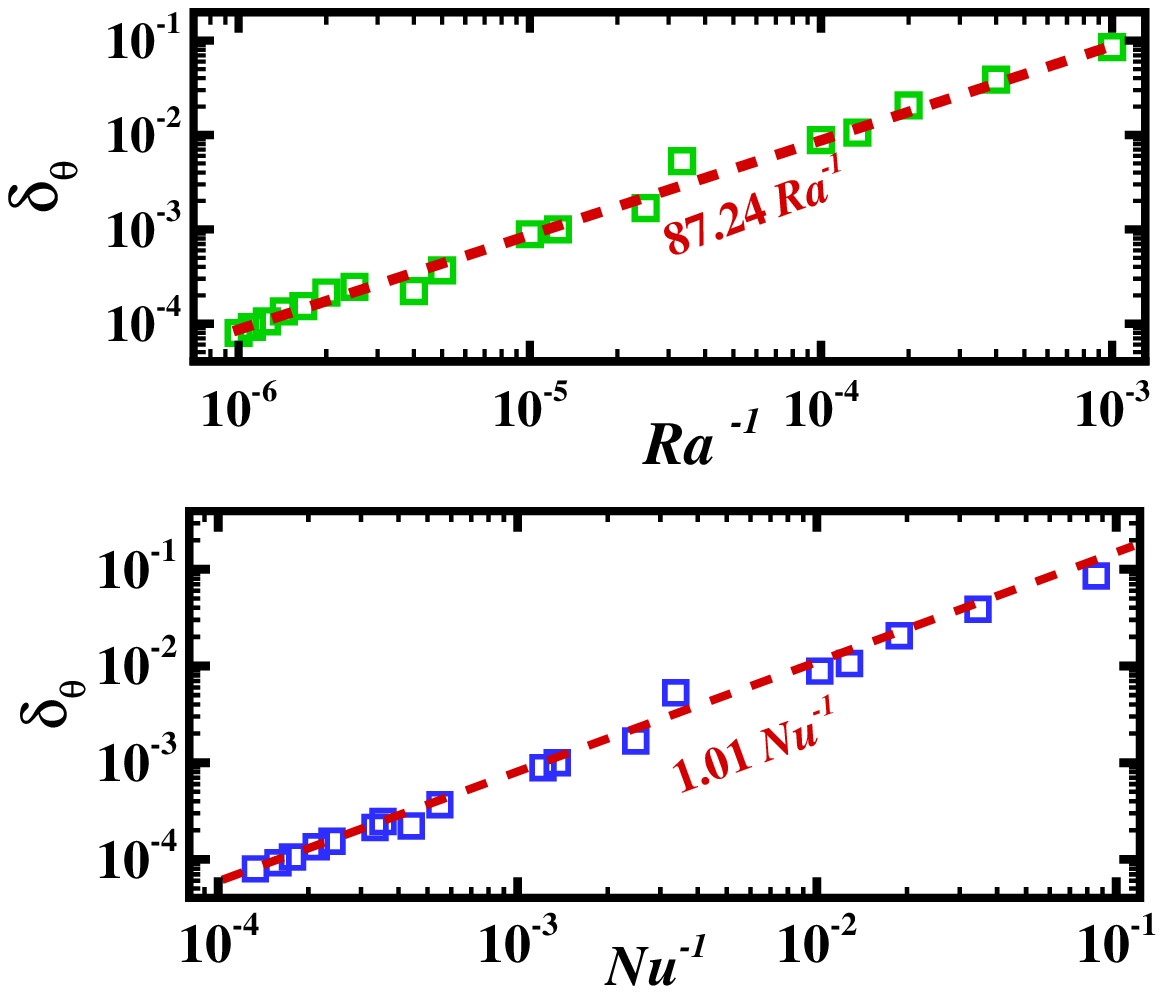}
        \caption{\label{fig:6b}}
    \end{subfigure}
 \captionsetup{justification=justified, singlelinecheck=false,width=\textwidth}

    \caption{\label{fig:6} \textit{(a)} Time- and horizontally averaged variance of temperature, \textit{(b)} variation of the  the thermal boundary layer thickness ($\delta_{\theta}$) with $Ra$ and $Nu$.}
  
\end{figure}

\begin{figure}
    \begin{subfigure}{1.0\linewidth}
         \includegraphics[width=\linewidth]{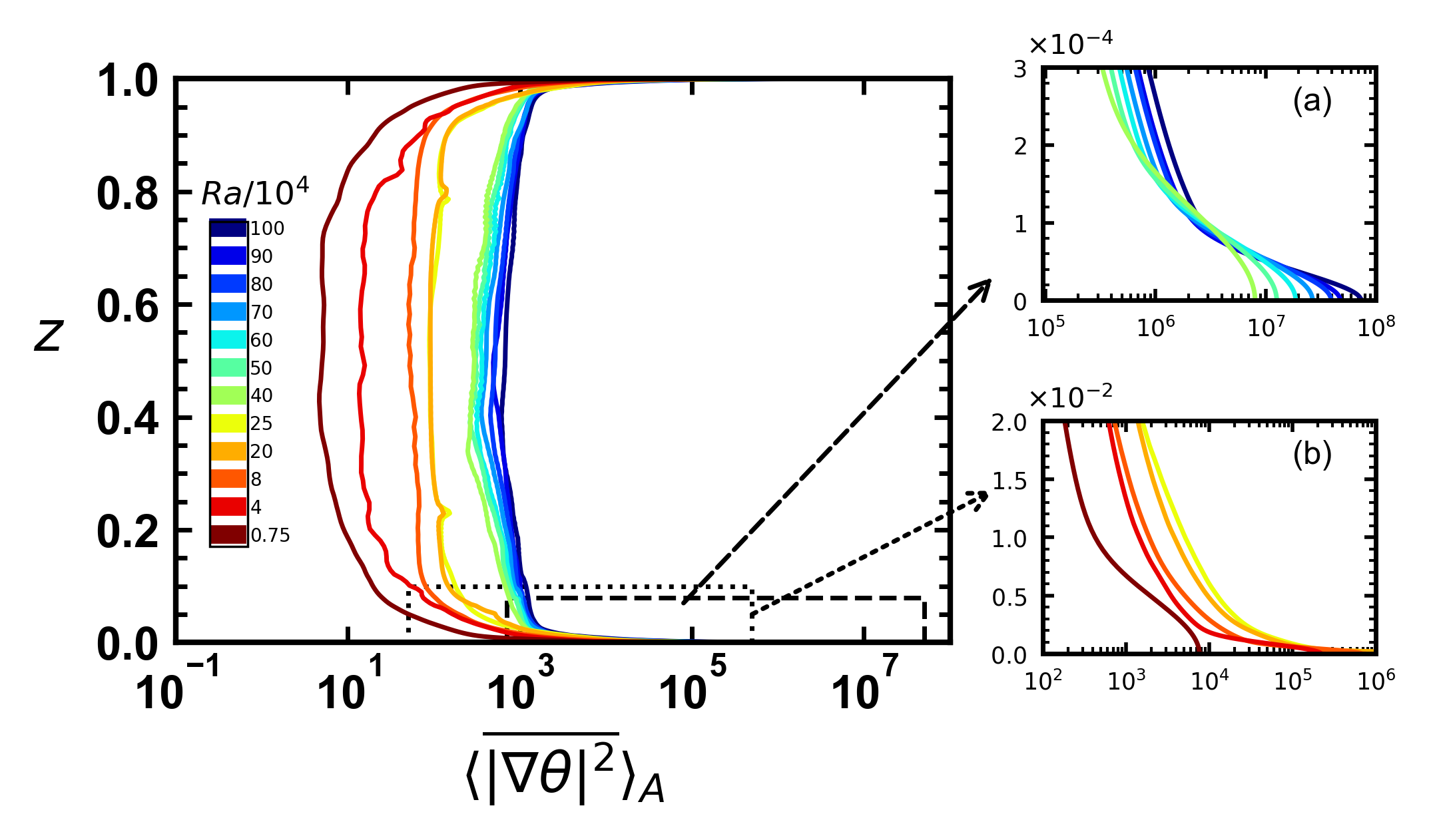}
        \caption{\label{fig:7a}}
    \end{subfigure}
    \hfill
    \begin{subfigure}{0.48\linewidth}
         \includegraphics[width=\linewidth]{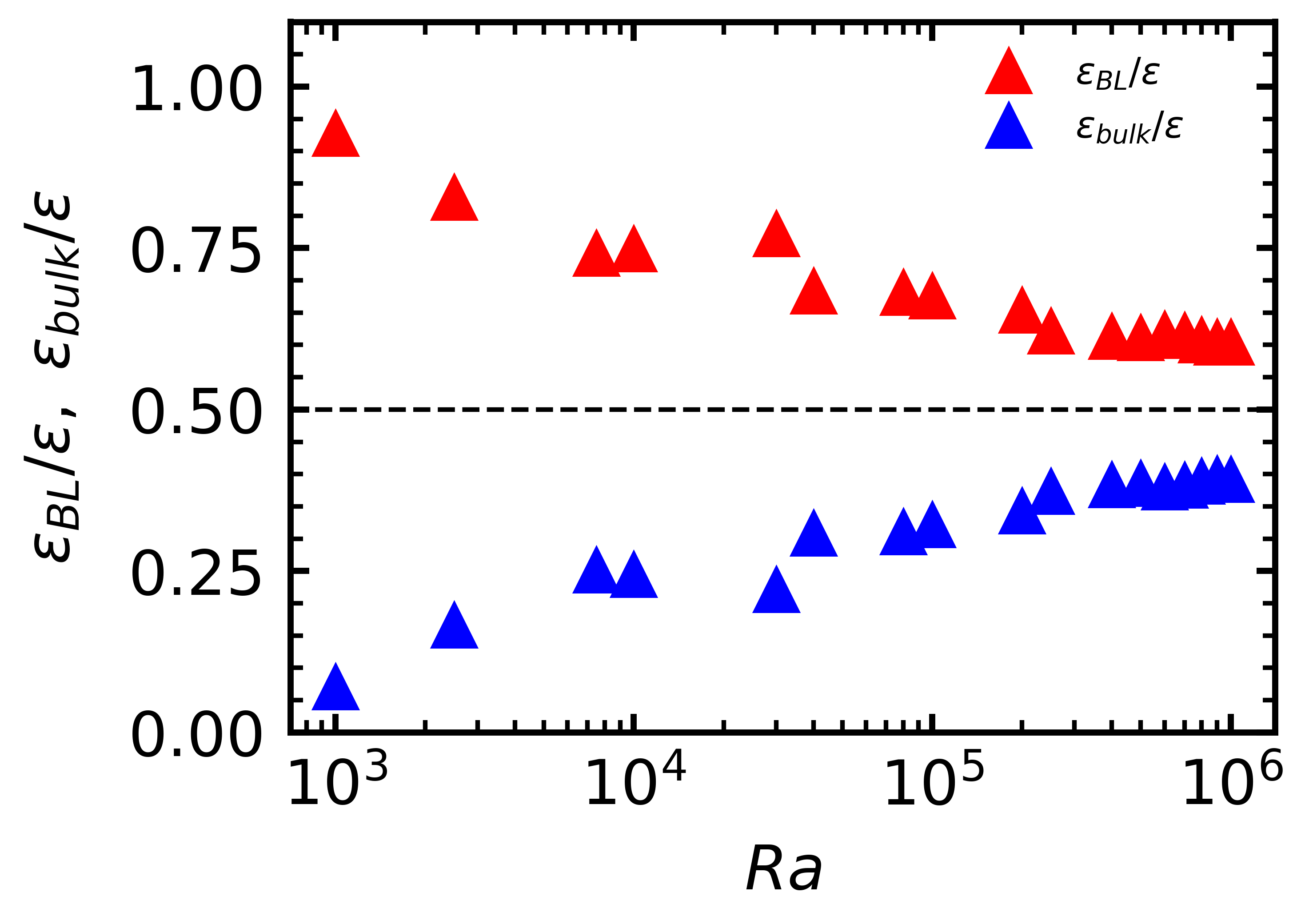}
        \caption{\label{fig:7b}}
    \end{subfigure}
 \captionsetup{justification=justified, singlelinecheck=false,width=\textwidth}
    \hfill
    \begin{subfigure}{0.49\linewidth}
         \includegraphics[width=\linewidth]{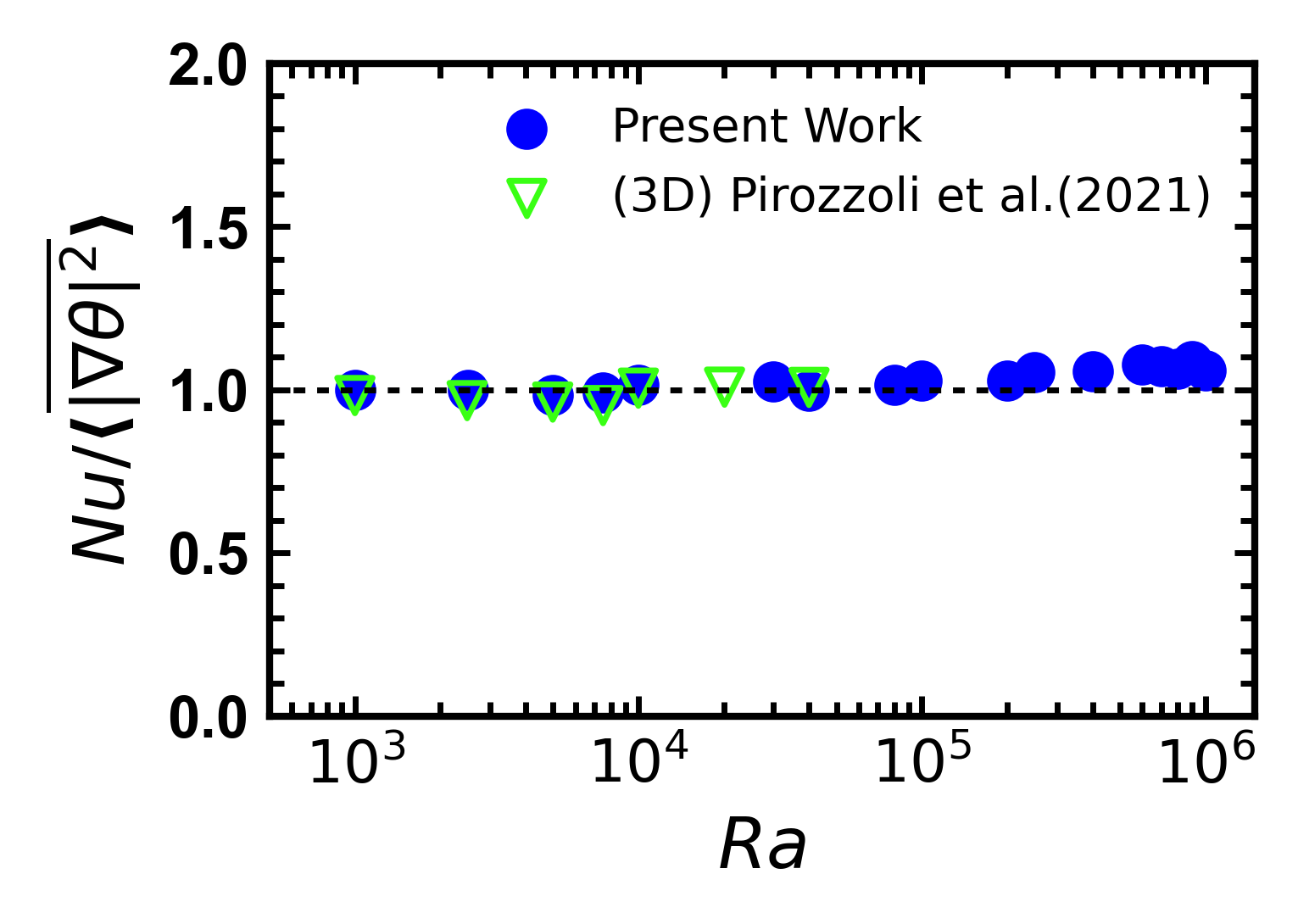}
        \caption{\label{fig:7c}}
    \end{subfigure}
    
    \caption{\label{fig:7} (a) Horizontal and time-averaged thermal dissipation profile
    across the vertical direction for different $Ra$,
     (b) Boundary-layer (BL) and bulk contributions to the mean scalar dissipation ($\epsilon$) for different cases. (c) Ratio of the Nusselt number to the dimensionless thermal dissipation for different $Ra$.}
\end{figure}

To further investigate the flow dynamics within the domain, we present the distribution of the horizontal and time-averaged temperature profiles along the vertical direction in figure \ref{fig:5}. The temperature profiles change from near boundaries towards the bulk of the domain. Approximate linear profiles in the center of the domain could be observed for all the cases, similar to the prediction of the three-dimensional 'heat-exchanger' solution of \cite{hewitt2014high}. We did not include the low $Ra$ cases ($1\times10^3, 2.5 \times 10^3$, $5 \times 10^3$, $3\times10^4$ and  $4\times10^4$) in figure \ref{fig:5} to improve readability. The inset (a) of figure \ref{fig:5} illustrates the calculation of the boundary layer thickness similar to  \cite{zhu2024transport} to capture the transition between the bulk and the boundary layer region for $Ra=2\times 10^5$. The boundary layer thickness is determined as the distance from the wall to the red marker, which represents the intersection point of the tangents to the near-wall temperature profile and the bulk temperature profile. We can see that the bulk temperature profile shifts towards $\langle \overline{\theta} \rangle_A = 0.5$ with increasing $Ra$. This indicates a steeper transition to the near-wall temperature profile, further implying a reduction in the boundary layer thickness with increasing $Ra$. \cite{hewitt2014high} reported similar observations for $Ra \leq 2 \times 10^4$. \\

To analyze the temperature evolution in the ultimate regime, the near-boundary temperature profiles are magnified in the inset (b) of figure \ref{fig:5} for $Ra\geq 1\times 10^5$. The transition of the temperature profile from the region near the wall to the bulk occurs earlier for $Ra \geq 2\times 10^5$ compared to $Ra= 1\times 10^5$. The near-wall temperature profile shrinks towards the wall for $Ra \geq 4 \times 10^5$ and reduces further with increasing $Ra$, indicating significant decrease in the boundary-layer thickness compared to $Ra \leq 1\times10^5$.\\

\cite{de2022strong} quantified the boundary layer thickness as the region where the temperature fluctuations $(\theta_{rms})$ increase sharply until they develop a peak. We plot the vertical variation of temperature variance $\overline{\langle \theta^{\prime 2} \rangle}$ for different cases in figure \ref{fig:6} (a). The peak location in the vertical direction marks the boundary layer thickness. After the peak, the $\overline{\langle \theta^{\prime 2} \rangle}$ gradually decreases towards the center of the domain. Table \ref{tab:case details} lists the boundary layer thickness ($\delta_{\theta}$) for all the cases. Notice that the boundary layer thickness for $Ra = 1\times 10^3$ and $8 \times 10^4$ are $\sim 10^{-1}$ and $\sim 10^{-3}$ respectively, similar to the results of \cite{de2022strong}. We also plot $\delta_{\theta}$ against $Ra$ in figure \ref{fig:6}(b) and find a linear scaling between $\delta_{\theta}$ and $Ra^{-1}$ consistent with the theoretical prediction with a coefficient of $87.24$. Similarly, we also find a linear scaling between $\delta_{\theta}$ and $Nu^{-1}$ with a coefficient consistent with the classical scaling $Nu \sim \delta{\color{black}_{\theta}}^{-1}$. Despite structural changes in the boundary layer, the observed scaling of the peak variance location with $Ra$ confirms that the inverse relationship between $Nu$ and $\delta_{\theta}$ also holds in the ultimate regime, supporting the generality of this scaling behavior. \\

To gain more insight into the thermal boundary layer dynamics, we present the vertical non-dimensional thermal dissipation profiles \citep{zhu2024transport} $\langle \overline{|\nabla \theta|^2} \rangle_A$, for different Rayleigh numbers in figure~\ref{fig:7a}. The enlarged inset (a) shows the near-wall thermal dissipation for $Ra > 2.5 \times 10^5$, whereas inset (b) shows the near-wall thermal dissipation profiles up to $Ra \leq 2.5 \times 10^5$. We can observe that with the increase in $Ra$, the near-wall thermal dissipation increases. This is attributed to the increase in the number and decrease in the size of the protoplumes emerging from the wall with increasing $Ra$, resulting in enhanced heat transport from the wall. We further compute the thermal dissipation within the boundary layer ($\epsilon_{BL})$ and in the bulk $(\epsilon_{bulk}$) for different cases and plot their ratio with respect to the total dissipation $(\epsilon)$ in figure \ref{fig:7b} to understand the efficacy of the protoplumes in transferring the heat from the wall to the interior. Interestingly, the ratio $\epsilon_{BL}/\epsilon$ decreases, and $\epsilon_{bulk}/\epsilon$ increases with increasing $Ra$, signifying the increase in the bulk thermal dissipation with increasing $Ra$. This observation is consistent with the fact that the smaller and increased numbers of protoplumes at higher $Ra$ efficiently transfer the heat from the wall towards the interior of the domain.    \\

\cite{zhu2024transport} correlated the thermal and the kinetic response parameters with the governing parameter $Ra$ by performing a volume and time average of the steady-state energy equation (Equation~\ref{Eq:energy_nondim}). They found that the ratio of the Nusselt number to the domain-averaged thermal dissipation must equal unity. It also signifies the conservation of thermal energy. We plot this ratio in figure \ref{fig:7b} and achieve values $\sim 1$, indicating sufficient grid resolution for conservation of thermal energy for all the cases. We also include the data points reported by \cite{zhu2024transport} for the cases from \citet{pirozzoli2021towards} for comparison.   \\

\subsection{Flow structures}
\label{sec:section3.4}

From the previous discussion, it is apparent that the flow structures near the wall influence the overall heat transfer. Therefore, it becomes imperative to quantify these flow structures near the wall and correlate their evolution with the overall heat transfer mechanism. For the quantitative estimation of flow structures and their associated heat transfer characteristics, we compute the two-dimensional spectral density $P(k_x,k_y)$ by taking the Fourier transform of the temperature field $\theta(x,y)$ at $z=30/Ra$, and at the midplane ($z=0.5$) \citep{otero2004high, pirozzoli2021towards,de2022strong}. Here, $k_x$ and $k_y$ denote the time-averaged wavenumbers in the respective directions.  

\begin{equation}\label{Eq:kr_mean}
\overline{k}(z)
=\Biggl \langle \frac{\int \int \sqrt{k_x^2 +k_y^2} P(k_x,k_y) dk_x dk_y}{\int \int P(k_x,k_y) dk_x dk_y } \Biggl \rangle 
\end{equation}

We further use $P(k_x, k_y)$ to compute the dominant mean wavenumber, $\overline{k}$, inverse size of the dominant structure, at $z=30/Ra$, and at the midplane ($z=0.5$) using equation \ref{Eq:kr_mean} similar to \cite{hewitt2014high,de2022strong}. Figure \ref{fig:9} displays the variation of $\overline{k}$ as a function of $Ra$ for different vertical locations. We find that the $\overline{k}$ scales linearly with $Ra$ and the data points could be fit using equation \ref{Eq:k_scaling_pp} in figure \ref{fig:9} (a), with 98.5\% confidence interval for $z=30/Ra$ for $Ra\leq 2.5\times 10^5$.
The observed increase in $\overline{k}|_{pp}$ with $Ra$ indicates that higher buoyancy forcing leads to finer plume spacing and smaller dominant horizontal scales, consistent with the transition toward more vigorous and fragmented convective dynamics.\\

\begin{figure}
         \begin{subfigure}{0.5\linewidth}
         \includegraphics[width=\linewidth]{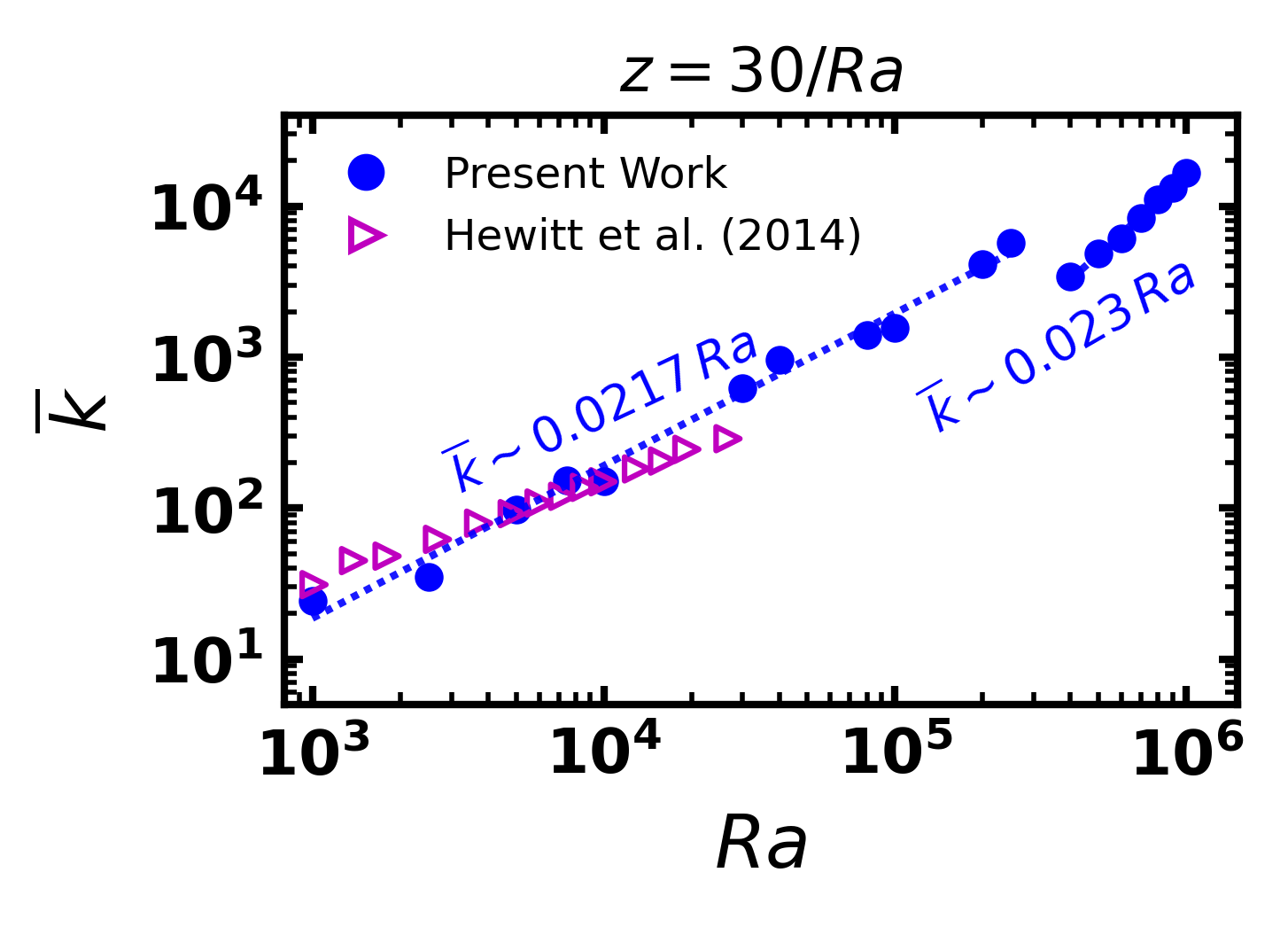}
        \caption{\label{fig:9}}
     \end{subfigure}
      \hfill
       \begin{subfigure}{0.5\linewidth}
         \includegraphics[width=\linewidth]{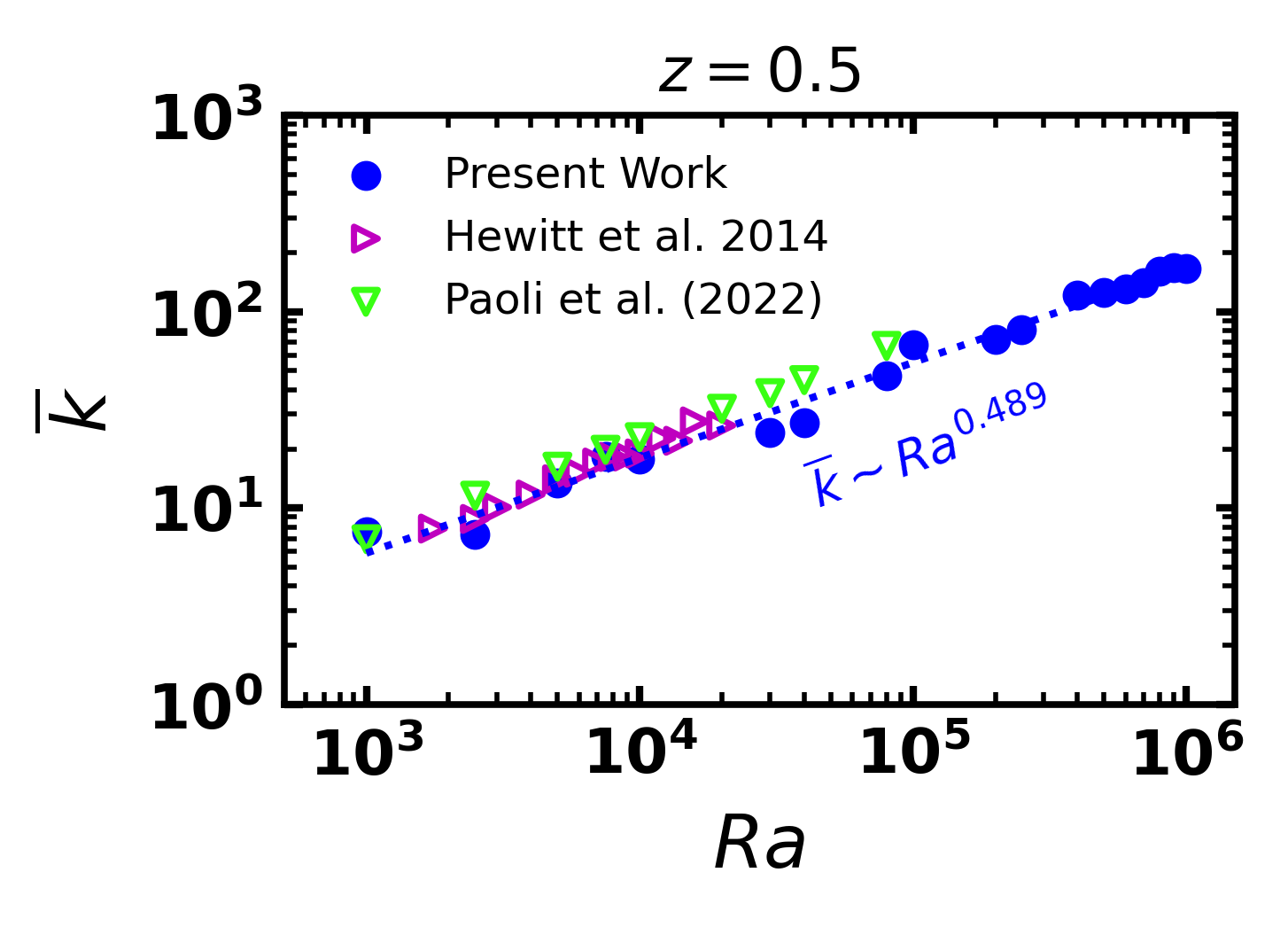}
         \caption{\label{fig:9}}
     \end{subfigure}
    
  
 \captionsetup{justification=justified, singlelinecheck=false,width=\textwidth}
    \caption{\label{fig:9}  The mean wavenumber of the temperature distribution is calculated using Equation \ref{Eq:kr_mean} for (a) near the wall (z=$30/Ra$) and (b) at the centre as a function of $Ra$ along with the comparison with available literature \cite{hewitt2014high}, \cite{de2022strong} for the respective locations. The best fits are $\overline{k}\sim Ra$ 
        and $\overline{k}\sim Ra^{0.489}$ respectively.}
\end{figure}

\begin{equation} \label{Eq:k_scaling_pp}
\overline{k}|_{pp}= \overline{k}(z=30/Ra)= 0.0217Ra-83.68\\
\end{equation}

For $Ra\geq 4\times10^5$ also, we find a linear variation of $\overline{k}$ with $Ra$. However, the data fit shows a different scaling than that obtained for $Ra \leq 2.5 \times 10^5$, which is represented by equation \ref{Eq:k_scaling_pp_ultimate} for $z=30/Ra$. \cite{hewitt2014high} reported the power law scaling $\overline{k} \sim Nu$ in the protoplume region ($z = 30/Ra$), which implies $\overline{k}\sim Ra$ as $Ra \rightarrow \infty$. In contrast, \cite{de2022strong} estimated $\overline{k}\sim Ra^{0.81}$ for $10^3 \leq Ra \leq 8\times 10^4$. They argued that such a sublinear exponent is reasonable in the early ultimate regime, likely around $Ra \approx 5\times 10^5$ suggested by \cite{pirozzoli2021towards}). Our measurements of the mean wavenumber ($\overline{k}$) versus $Ra$ agree with the linear scaling reported by \cite{hewitt2014high} across both the early and ultimate regimes. This linear relation indicates that the horizontal size of near-wall plumes scales directly with the boundary-layer thickness.\\

 \begin{equation} \label{Eq:k_scaling_pp_ultimate}
 \overline{k}|_{pp}= \overline{k}(z=30/Ra)= 0.023Ra - 6897.7 \\
 \end{equation}
For the midplane, \cite{hewitt2014high} numerically reported a scaling of $\overline{k}\sim Ra^{0.52}$ for $1750\leq Ra \leq 2\times10^4$ and highlighted the need to validate the asymptotic scaling at higher $Ra$. Consistent with this, \cite{de2022strong} reported $\overline{k}\sim Ra^{0.49}$ for $Ra\leq 8\times 10^4$. A theoretical framework developed by \cite{hewitt2017stability}, based on a heat-exchanger analogy, further explained that 
the vigrous mixing of the protoplumes ($\overline{k} \sim Ra$) at the boundaries continually force the interior flow, which must coarsen to at least $\overline{k}\sim Ra^{0.5}$ to sustain heat transport across the domain. Our data fit (equation \ref{Eq:k_scaling_mid} for $Ra \leq 1\times10^6$) with a 98.9\% confidence interval and confirms this prediction (shown in figure \ref{fig:9}(b)), demonstrating close agreement with both numerical and theoretical estimates for higher $Ra$. 
\begin{equation} \label{Eq:k_scaling_mid}
\overline{k}(z=1/2)=0.201Ra^{0.489}
\end{equation}
This agreement indicates that the flow exhibits weaker scaling in the interior compared to the near-wall regions.\\

\section{Conclusion} \label{section4}
We perform DNS to explore Rayleigh-Darcy convection at Rayleigh numbers ranging from $10^3$ to $10^6$ in a three-dimensional domain to get an estimate of the $Nu$ and report a relation between the $Nu$ and $Ra$ for the unexplored ultimate regime at $Ra \geq 5 \times 10^5$ relevant to the real-world geological situations, such as the Utsira Sand reservoir at Sleipner. Additionally, we aim to quantify the length scale associated with the near-wall flow field.  We found a linear variation of the $Nu$ with $Ra$ for $1\times 10^3 \leq Ra \leq 1\times 10^6$. However, we report a transition in this linear variation from $Ra \sim 4 \times 10^5$, indicating the beginning of the ultimate regime in Rayleigh-Darcy convection. We linear fit our dataset and report that our scaling for $Ra \leq 2.5 \times 10^5$ is $\sim 6.25\%$ lesser than the scaling of \cite{hewitt2014high}. 
Additionally, for $Ra \geq 4 \times 10^5$, our $Nu$ values are $\sim1.24 \%$ lower than the extrapolated prediction of \cite{pirozzoli2021towards} for the ultimate regime ($Nu=0.0081Ra$ for $Ra\geq 5\times 10^5$), signifying a close agreement with the literature. To further corroborate this transition, we plot the compensated Nusselt number and found approximately constant values similar to those suggested by \cite{pirozzoli2021towards}. Similar transition in the compensated Sherwood number for $\sim Ra >2\times10^4$ was reported from the quasi-2D numerical ($Ra \leq 4 \times10^4$) and experimental ($Ra \sim 10^6 $) datasets of \cite{neufeld2010convective}. \\

We observed small-scale structures: protoplumes near the boundaries, which converge toward the centerline and span across the domain as large columnar structures: megaplumes.  With the increase in $Ra$, the number of protoplumes increases, indicating enhanced convection at the boundaries and efficient heat transport. We further quantified the thermal boundary layer thickness ($\delta_{\theta}$) as the variance
of the temperature field. We found that despite the structural variability of the thermal plumes near the wall in the ultimate regime,   $\delta_{\theta}$ varies as the inverse of $Ra$. Similarly, our data also demonstrates that $\delta_{\theta} \sim Nu^{-1}$, thus indicating the validity of the linear variation for the ultimate regime and corroborating the generality of this scaling behavior. To further understand the thermal boundary layer dynamics, we compute the thermal dissipation within the boundary layer and within the bulk. We found that the bulk dissipation increases with increasing $Ra$. In contrast, the dissipation within the boundary layer decreases, signifying that the protoplumes efficiently transfer heat from the walls to the interior. We also compared $Nu$ with the volume-averaged thermal dissipation, and found their ratio $\sim 1$ for all the cases, signifying that our grid resolution in all the directions is sufficient to capture small and large-scale dynamics accurately. Note that for the high $Ra$ cases, our horizontal domain is $L_x = L_y = 0.015625$, while the vertical domain size $L_z$ is kept constant at unity for all the cases. This selection of lower horizontal domain size is motivated by the investigation of \cite{iyer2020classical}, where it was reported that the temperature and the fluctuating velocity fields are similar for aspect ratios of $0.1$ and $1$ owing to a very thin boundary layer compared to the horizontal dimensions of the domain at high $Ra$. Additionally, the near-wall dominant wavenumber calculation reveals that the domain length for all our cases is able to well accommodate the largest wavelengths, thus corroborating the use of lower aspect ratio domains for the high $Ra$ cases. The thermal gradients become increasingly concentrated near the walls as $Ra$ increases. Therefore, we have used grid clustering in the vertical direction and ensured enough grid points within $\delta_{\theta}$ to resolve the thermal gradients accurately. \\

Finally, we assess the flow structures to understand the overall heat transfer mechanism. We compute the dominant length scale using the temperature field near the wall ($30/Ra$) and at the mid-plane $(z = 0.5)$. Near the wall, the mean wavenumber linearly varies with $Ra$. Noticeably, the slope of this linear variation is higher in the ultimate regime, $Ra \geq 4 \times 10^5$, than that for $Ra \leq 2.5 \times 10^5$, signifying a decrease in the size of the protoplumes. Owing to the decrease in size and increase in the number of these protoplumes in the ultimate regime, heat transfer enhances. The linear scaling of the mean wavenumber with $Ra$ changes to $~Ra^{0.489}$ consistent with the findings of \cite{hewitt2014high}, signifying the merging of the protoplumes to form megaplumes. This power scaling indicates that the megaplumes also become finer with increasing $Ra$ in the ultimate regime, thus leading to efficient heat transport in the bulk. \\

\bibliographystyle{jfm}
\bibliography{jfm}








\end{document}